\documentclass[aps,prb,reprint,10pt,superscriptaddress,amsmath,amssymb,floatfix]{revtex4-2}
\usepackage[T1]{fontenc}
\usepackage{newtxtext,newtxmath}
\usepackage{booktabs}
\usepackage{graphicx}
\usepackage{tikz}
\usepackage{hyperref}
\usetikzlibrary{arrows.meta,calc,positioning}
\begin{document}
\title{An Efficient Method for Gibbs Free Energy Evaluation under Volume Compression}

\author{Zhiyuan Gao}
\affiliation{Key Laboratory of Photovoltaic and Energy Conservation Materials, Institute of Solid State Physics, HFIPS, Chinese Academy of Sciences, Hefei 230031, China}
\affiliation{Science Island Branch of Graduate School, University of Science and Technology of China, Hefei 230026, China}

\author{Yong Yang}
\email{yyanglab@issp.ac.cn}
\affiliation{Key Laboratory of Photovoltaic and Energy Conservation Materials, Institute of Solid State Physics, HFIPS, Chinese Academy of Sciences, Hefei 230031, China}
\affiliation{Science Island Branch of Graduate School, University of Science and Technology of China, Hefei 230026, China}

\author{Yoshiyuki Kawazoe}
\affiliation{New Industry Creation Hatchery Center, Tohoku University, Sendai, 980-8579, Japan}
\affiliation{Center for Interdisciplinary Research, SRM University-AP, Neerukonda, Mangalagiri Mandal, Guntur District, Andhra Pradesh, 522240, India}

\begin{abstract}
Accurate evaluation of Gibbs free energies is essential for constructing
pressure-temperature phase diagrams. Conventional methods based on the
quasi-harmonic approximation (QHA) require phonon spectra at many volume
points and are therefore expensive in general. Here we develop an efficient
method based on the interpolation of a few \emph{ab initio} data points for Gibbs
free energy evaluation under volume compression. Phonon spectra are calculated
only at selected volumes. An effective Gr\"uneisen parameter derived
from the zero-point energy (ZPE) reconstructs the static--ZPE branch, while
piecewise mode-resolved Gr\"uneisen slopes reconstruct the finite-temperature
vibrational branches on the target volume grids. The method is validated
against QHA benchmarks for diamond (C), Al, Si, Ge, rutile TiO$_2$,
$\beta$-PtO$_2$, and
Ta$_2$O$_5$ polymorphs. For simple benchmark systems (C, Al, Si, Ge,
rutile TiO$_2$, and $\beta$-PtO$_2$), the Gibbs free energy mean absolute errors (MAEs)
relative to the QHA benchmarks remain below 0.53~meV/atom, with a six-system
average of 0.148~meV/atom, while
the number of explicit phonon volume points is reduced from about 20--21 to
3 in the lowest-cost implementation. For the more complex Ta$_2$O$_5$
polymorphs, the reconstructed free energies reproduce the main
phase-stability topology despite larger phase-dependent errors. With reference
to the QHA workflows, the interpolation method in this work achieves speedups
of 5.911--9.023$\times$ and remains reliable for moderate compression ranges
where phonon frequencies vary smoothly with volume.
\end{abstract}
\maketitle

\section{Introduction}

Recent advances in artificial intelligence (AI) and high-throughput methods
have greatly expanded the chemical and structural search space available for
materials design. For instance, graph-network-based materials exploration has
predicted millions of crystal structures, universal graph interatomic
potentials such as M3GNet and CHGNet have accelerated zero-temperature
structural relaxation and energy screening over large chemical spaces, and
generative models such as MatterGen show that inorganic crystals can be
generated under target property
constraints\cite{merchant2023gnome,chen2022m3gnet,deng2023chgnet,zeni2025mattergen}.
These developments have substantially changed the initial stage of materials
discovery: candidate generation and ground-state ranking can now be performed
at a much larger scale than was previously accessible. The bottleneck
therefore shifts from producing candidate structures to verifying their
thermodynamic stability under the finite-temperature and finite-pressure
conditions relevant to synthesis and application.

In this context, thermodynamic quantities, in particular Gibbs free energies,
provide the downstream validation needed after zero-temperature (0~K)
prescreening. Many large-scale screens based on numerical calculations
initially prioritize structures by 0~K energies or by the energy above the
convex hull\cite{jain2013materialsproject,merchant2023gnome,chen2022m3gnet}.
Such criteria are essential for reducing the search space, but they do not
determine which polymorph is stable at finite temperature and pressure.
Zero-point motion, vibrational entropy, and pressure-volume work can reorder
competing phases and shift phase boundaries. Consequently, phase-stability
studies compare the Gibbs free energy as a function of temperature $T$ and
pressure $P$, $G(T,P)$, rather than static energies
alone\cite{wallace1972,grabowski2007,togo2015,lee2013znsno}. For experimental
translation, the relevant question is not only whether an AI-generated
candidate has a low 0~K energy, but whether it occupies an accessible
temperature-pressure stability field. Finite-temperature and finite-pressure
phase diagrams therefore connect candidate discovery to synthesis-relevant
stability windows and phase-boundary predictions\cite{dinsdale1991,liu2023}.
For nonmagnetic semiconductors and insulators, where electronic excitations
are usually weak because of a finite band gap, lattice vibrational free
energies often provide the dominant finite-temperature contribution
\cite{grabowski2007,dove1993,togo2015}.

Therefore, accurate calculation of $G(T,P)$ is central to the description of
phase stability, phase transformation, and other thermodynamic properties of
materials\cite{chen2002,steinbach2009,liu2023}. Although experimental data
have enabled the construction of thermodynamic databases within the
Calculation of Phase Diagrams (CALPHAD) framework\cite{dinsdale1991,liu2023},
first-principles calculations based on density functional theory
(DFT)\cite{martin2004} are increasingly needed to complement missing data and
assess materials under conditions that are difficult to access
experimentally.

The difficulty is that first-principles evaluation of volume-dependent
vibrational free energies remains computationally expensive. Density
functional perturbation theory (DFPT) phonon calculations require accurate
self-consistent perturbation calculations and dense sampling of the phonon
spectrum\cite{baroni2001,togo2015}. In conventional dense-volume
quasi-harmonic approximation (QHA)
workflows, these phonon-related quantities are usually evaluated at many
volume points, typically 10--20, before constructing the Gibbs free energy
surface\cite{wallace1972,grabowski2007,togo2015}. This dense volume sampling
is the computational bottleneck between fast AI-assisted
candidate generation and rigorous finite-temperature phase-stability
verification. For structurally complex materials such as the Ta$_2$O$_5$
polymorphs examined in this work, whose primitive cells contain
$Z=1$--11 formula units and therefore 7--77 atoms, repeated
volume-dependent phonon calculations with larger supercells become especially
costly for Gibbs free energy calculation and phase-diagram construction.

Machine-learning interatomic potentials can accelerate structural
relaxation, molecular and lattice-dynamics calculations, and have achieved
near-DFT accuracy in selected systems\cite{ladygin2020lattice}. However, for
chemically unusual or strongly anharmonic systems, reliable use still
requires system-specific training or fine-tuning data. The volume integral of
pressure (VIP) method developed by Hashimoto \emph{et al.} provides another
efficient approach for free energy evaluation under thermal
expansion\cite{hashimoto2025}. VIP targets the small-strain thermal-expansion
regime, typically of the order of $10^{-4}$ to $10^{-3}$, and is based on a
mathematical expansion of the free energy around a reference volume point. A
key approximation employed by VIP is that the volume dependence of the product
between the Gr\"uneisen parameter and phonon frequency is negligible. These
approaches therefore address complementary regimes of the broader
finite-temperature free energy problem.

In this work, we develop a method that enables efficient and accurate
evaluation of Gibbs free energies under significant volume compression and
expansion, with the relative magnitude of volume change $\Delta V/V$ being of
the order of $0.1$ for compression and $0.01$ for thermal expansion. The
method is motivated by the observation that, within a moderate range of
compression or expansion, phonon frequencies vary smoothly with the
logarithmic volume coordinate. It uses sparse-volume Gr\"uneisen interpolation (GI)
as the primary reconstruction route. The ZPE-level Gr\"uneisen
parameter reconstructs the static--ZPE branch, while piecewise mode-resolved
Gr\"uneisen parameters reconstruct the finite-temperature vibrational branch.
Using more sparse volumes generally improves the robustness of the local
GI, whereas using fewer points gives the largest
computational saving. The computational saving comes from avoiding explicit
phonon calculations at every volume needed for the final free energy surface.
In this sense, the method does not replace structure generation or 0~K
relaxation; instead, it reduces the cost of the finite-temperature Gibbs free
energy check that follows candidate generation.

The GI methodology is validated through systematic
comparisons with QHA data for typical benchmark systems including C, Al, Si,
Ge, rutile TiO$_2$, $\beta$-PtO$_2$, and Ta$_2$O$_5$ polymorphs. It is
demonstrated that the method reproduces QHA Gibbs free energies with small
errors for the nonmagnetic crystalline systems investigated here while
substantially reducing the number of explicit phonon calculations. The extracted
Gr\"uneisen parameters also provide a compact measure of volume sensitivity of
phonons: phases with weaker frequency changes under compression give smaller
extracted Gr\"uneisen parameter ($\gamma$) values, whereas phases with stronger
volume-sensitive modes give larger values. Within this tested domain, the
proposed GI method serves as a practically efficient
method for first-principles phase-diagram construction involving
volume-dependent thermodynamic quantities.

The remainder of this article is organized as follows. Section~II presents
the formulation of reduced-volume GI, including the
ZPE-level scaling for the static--ZPE branch, the piecewise mode-resolved
GI for the finite-temperature vibrational branch, and
the details of first-principles calculations. Section~III first
validates the method on elemental and binary benchmark systems, then applies
it to the competing Ta$_2$O$_5$ polymorphs and pressure-temperature phase
diagrams. Furthermore, the same section examines thermal-expansion
coefficients for Al and Si, compares the time cost with QHA calculations, and
discusses the applicable compression range of the method. Section~IV
summarizes the main conclusions and the scope of the proposed workflow.

\section{Theory}

\subsection{Mode-Resolved GI of Vibrational Frequency}

The QHA calculations used for benchmarking evaluate phonon spectra on a dense
volume grid. The reduced-volume workflow used here
calculates phonons only at a sparse set of $N_{\mathrm{sp}}$ stable volumes
$\{V_i\}_{i=1}^{N_{\mathrm{sp}}}$ and reconstructs the target volumes
$\{V_m\}$ through GI. Here, the indices $i$ and
$m$ label the sparse volumes used for the GI reconstruction and the target reconstructed volumes,
respectively, and $N_{\mathrm{sp}}$ is the number of sparse phonon-volume points.
At the sparse volumes, the phonon spectra $\omega_j(V_i)$, zero-point
energies (ZPEs) $\mathrm{ZPE}(V_i)$, and static DFT total energies $U(V_i)$
are obtained explicitly. Here, $U(V)$ is obtained at fixed volume and excludes
phonon contributions. The compact mode index $j$ represents
$(\vec q,n)$, where $\vec q$ denotes the phonon wave vector and $n$ is the
phonon branch index. Unless otherwise stated, $\sum_j$ denotes the
corresponding weighted sum over phonon wave vectors and branches with the
same normalization used in the QHA free energies.

Within QHA\cite{wallace1972,dove1993,togo2015}, the Gibbs free energy
$G(T,V;P)$ at temperature $T$, volume $V$, and pressure $P$ may be written as
\begin{align}
G(T,V;P)
&=
F(T,V)+PV,
\notag\\
F(T,V)
&=
U(V)+\mathrm{ZPE}(V)+F_v(T,V),
\notag\\
\mathrm{ZPE}(V)
&=
\frac{1}{2}\sum\nolimits_j \hbar\omega_j(V),
\notag\\
F_v(T,V)
&=
k_B T\sum\nolimits_j
\ln\!\left[
1-\exp\!\left(
-\frac{\hbar\omega_j(V)}{k_BT}
\right)
\right].
\label{eq:qha_free_energy}
\end{align}
Here, $F(T,V)$ is the Helmholtz free energy; $PV$ is the
pressure-volume work term; $\mathrm{ZPE}(V)$ is the zero-point energy (ZPE);
and $F_v(T,V)$ is the finite-temperature vibrational free energy contribution.
The constants $\hbar$ and $k_B$ denote the reduced Planck constant and
Boltzmann constant, respectively.
Equivalently, when the phonon density of states (PDOS) $g(\omega,V)$ is used, the
finite-temperature vibrational term can be evaluated as
\begin{equation}
F_v(T,V)
=
k_B T\int_0^{\omega_{\max}}
g(\omega,V)
\ln\!\left[
1-\exp\!\left(-\frac{\hbar\omega}{k_BT}\right)
\right]\mathrm d\omega .
\label{eq:pdos_thermal}
\end{equation}
Here, $g(\omega,V)$ is the PDOS at volume $V$, $\omega$ is
the phonon frequency, and $\omega_{\max}$ is the maximum frequency in the
sampled spectrum.
Numerically, the PDOS integral is evaluated on a discrete frequency grid. For
a grid spacing $\Delta\omega$,
\begin{align}
\int_0^{\omega_{\max}}g(\omega,V)\mathrm d\omega
&\simeq
\sum_\ell g(\omega_\ell,V)\Delta\omega
=3N,
\notag\\
F_v(T,V)
&\simeq
k_BT\sum_\ell g(\omega_\ell,V)\Delta\omega
\notag\\
&\quad\times
\ln\!\left[
1-\exp\!\left(
-\frac{\hbar\omega_\ell}{k_BT}
\right)
\right],
\label{eq:pdos_discrete}
\end{align}
where $N$ is the number of atoms in the simulation cell and $\ell$ labels the
discrete PDOS frequency bins. Thus, once the volume dependence of $\omega_j(V)$ is known, both the
zero-point and finite-temperature phonon terms can be reconstructed without
performing an explicit phonon calculation at every target volume.

For metallic Al, a small correction to the free energy due to electronic
excitations at $T>0$ was also included using the DOS/Fermi-Dirac formulation
adopted in the VIP work of
Hashimoto \emph{et al.}\cite{hashimoto2025} and the electronic free energy
method of Zhang \emph{et al.}\cite{zhang2017electronic}. In the present
notation, the vibrational contribution is already split into the zero-point and
finite-temperature phonon parts, and the metallic electronic contribution is
added as a relative correction,
\begin{equation}
F(T,V)=U(V)+\mathrm{ZPE}(V)+F_v(T,V)+\Delta F_{\mathrm{ele}}(T,V).
\label{eq:qha_with_electronic}
\end{equation}
Here $\Delta F_{\mathrm{ele}}$ is measured relative to the zero-temperature
electronic free energy at the same volume, which avoids double counting the
0~K electronic contribution already included in $U(V)$. The $T=0$ term below is
understood as the zero-temperature limit of the same DOS integral:
\begin{align}
\Delta F_{\mathrm{ele}}(T,V)
&=
F_{\mathrm{ele}}(T,V)-F_{\mathrm{ele}}(0,V),
\notag\\
F_{\mathrm{ele}}(T,V)
&=
E_{\mathrm{ele}}(T,V)-T S_{\mathrm{ele}}(T,V),
\notag\\
E_{\mathrm{ele}}(T,V)
&=
\int_{-\infty}^{\infty}
\epsilon D(V,\epsilon)f(T,\epsilon)\,\mathrm d\epsilon,
\notag\\
S_{\mathrm{ele}}(T,V)
&=
k_B\int_{-\infty}^{\infty}
D(V,\epsilon)s(T,\epsilon)\,\mathrm d\epsilon .
\label{eq:electronic_free_energy}
\end{align}
In Eq.~\ref{eq:electronic_free_energy}, $F_{\mathrm{ele}}$ denotes the
DOS-integrated electronic free energy before the zero-temperature subtraction;
only $\Delta F_{\mathrm{ele}}$ is added to the total Helmholtz free energy in
Eq.~\ref{eq:qha_with_electronic}. Here, $E_{\mathrm{ele}}$ is the corresponding
finite-temperature electronic internal-energy term, $S_{\mathrm{ele}}$ is the
electronic entropy, $D(V,\epsilon)$ is the electronic density of states (DOS) at
volume $V$ and electron energy $\epsilon$, and $s(T,\epsilon)$ is the entropy
contribution of a single electronic state. The single-state entropy and
Fermi-Dirac occupation are
\begin{align}
s(T,\epsilon)
&=
\begin{aligned}[t]
-\bigl[
&f(T,\epsilon)\ln f(T,\epsilon)
\\
&+\{1-f(T,\epsilon)\}
\ln\{1-f(T,\epsilon)\}
\bigr],
\end{aligned}
\notag\\
f(T,\epsilon)
&=
\left[
1+\exp\!\left(
\frac{\epsilon-\epsilon_F(T)}{k_BT}
\right)
\right]^{-1},
\label{eq:fermi_entropy}
\end{align}
where $\epsilon_F(T)$ is the Fermi level determined by charge conservation,
\begin{equation}
N_{\mathrm{ele}}
=
\int_{-\infty}^{\infty}
D(V,\epsilon)f(T,\epsilon)\,\mathrm d\epsilon .
\label{eq:electron_number}
\end{equation}
Here, $N_{\mathrm{ele}}$ is the total number of electrons included in the DOS
integration. For each selected volume, $\epsilon_F(T)$ is solved using the
corresponding $D(V,\epsilon)$; thus the volume dependence of the DOS is
intrinsically included in the integral. In the calculations of Al,
$D(V,\epsilon)$ was obtained from ground-state static DOS
calculations on the same volume grid. The DOS used for
the temperature-dependent electronic entropy was then interpolated along the
equilibrium thermal-expansion path, $D[V_{\mathrm{eq}}(T),\epsilon]$. Thus, the
reported Al electronic entropy includes the effect of thermal expansion through
the volume dependence of the static DOS, but it does not include additional
DOS broadening from thermally displaced atomic configurations or explicit
electron-phonon renormalization. This follows the fcc-Al treatment of
Ref.~\cite{hashimoto2025}, where the displaced-configuration DOS correction was
reported to be negligible for Al and was mainly used for Ti. For the electronic
free energy surface used in the QHA/GI comparison for Al, a low-temperature
intercept correction was applied to $\Delta F_{\mathrm{ele}}(T,V)$ so that
$\Delta F_{\mathrm{ele}}(0,V)=0$; this correction changes only the electronic
free energy column and does not modify the plotted $S_{\mathrm{ele}}(T)$.

Within a moderate volume interval, the leading frequency-volume response may
be described by the Gr\"uneisen relation\cite{gruneisen1912}
\begin{equation}
\gamma
=
-\frac{\mathrm d\ln\omega}{\mathrm d\ln V}
=
\frac{\mathrm d\ln(\omega/\omega_0)}
{\mathrm d\ln(V_0/V)} ,
\label{eq:gamma_definition}
\end{equation}
where $\gamma$ is the Gr\"uneisen parameter, $\omega$ is the phonon frequency,
$V_0$ is the reference volume, and $\omega_0$ is the corresponding reference
frequency. For the approximately linear part of
$\ln(\omega/\omega_0)$ versus $\ln(V_0/V)$, one has
\begin{align}
\ln\!\left(\frac{\omega}{\omega_0}\right)
&=
\gamma\ln\!\left(\frac{V_0}{V}\right),
\notag\\
\omega(V)
&=
\omega_0
\exp\!\left[
\gamma\ln\!\left(\frac{V_0}{V}\right)
\right]
\equiv
\lambda(V)\omega_0 .
\label{eq:gamma_scaling}
\end{align}
Here $\lambda(V)=\exp[\gamma\ln(V_0/V)]$ is the frequency scaling factor
describing the change of volume from $V_0$ to $V$.

For the static--ZPE branch, the effective Gr\"uneisen parameter is estimated
from the logarithmic ZPE ratio at the selected sparse volumes. For simplicity
we define
\begin{equation}
\eta(V)=\ln\!\left[\frac{\mathrm{ZPE}(V)}{\mathrm{ZPE}(V_0)}\right],
\label{eq:eta_zpe}
\end{equation}
then the through-origin slope gives
\begin{align}
\gamma_{\mathrm{ZPE}}
&=
\frac{
\sum_i
\ln\!\left(\frac{V_0}{V_i}\right)
\eta(V_i)
}{
\sum_i
\left[\ln\!\left(\frac{V_0}{V_i}\right)\right]^2
},
\notag\\
H_s^{\mathrm{GI}}(V_m)
&=
U(V_m)
+\mathrm{ZPE}(V_0)
\exp\!\left[
\gamma_{\mathrm{ZPE}}
\ln\!\left(\frac{V_0}{V_m}\right)
\right].
\label{eq:static_zpe_gi}
\end{align}
Here, $i$ indexes the selected sparse volumes, $m$ indexes the target
volumes on the reconstructed grid, $\gamma_{\mathrm{ZPE}}$ is the ZPE-level
effective Gr\"uneisen parameter, and $H_s^{\mathrm{GI}}$ is the
GI static--ZPE contribution.
This step keeps the same ZPE-based Gr\"uneisen
scaling: the exponential factor reconstructs the ZPE at
the target volume, while $U(V_m)$ is taken from the static equation-of-state
volume grid or from the corresponding static DFT calculation.

For the finite-temperature vibrational branches, using a single scaling factor
for all modes can be oversimplified. We therefore keep the Gr\"uneisen form
but make the slope local in both phonon mode and volume interval. If the
target volume $V_m$ lies between two neighboring sparse points $V_a$ and
$V_b$ in the logarithmic volume coordinate, then
\begin{align}
\gamma_j^{ab}
&=
\frac{
\ln[\omega_j(V_b)/\omega_j(V_a)]
}{
\ln(V_0/V_b)-\ln(V_0/V_a)
},
\notag\\
\omega_j^{\mathrm{GI}}(V_m)
&=
\omega_j(V_a)
\exp\!\left\{
\gamma_j^{ab}
\left[
\ln\!\left(\frac{V_0}{V_m}\right)
-\ln\!\left(\frac{V_0}{V_a}\right)
\right]
\right\}.
\label{eq:mode_resolved_gi}
\end{align}
Here, $a$ and $b$ label the two sparse-volume endpoints that bracket the target
volume $V_m$, $\gamma_j^{ab}$ is the local mode-resolved Gr\"uneisen slope for
mode $j$ on that interval, and $\omega_j^{\mathrm{GI}}(V_m)$ is the
reconstructed mode frequency at $V_m$.
The same $\vec q$-point mesh and branch index are kept unchanged so that the
mode $j$ is followed consistently across the sparse-volume grid. The local
slopes $\gamma_j^{ab}$ are extracted from dynamically stable sparse spectra
with positive endpoint frequencies.

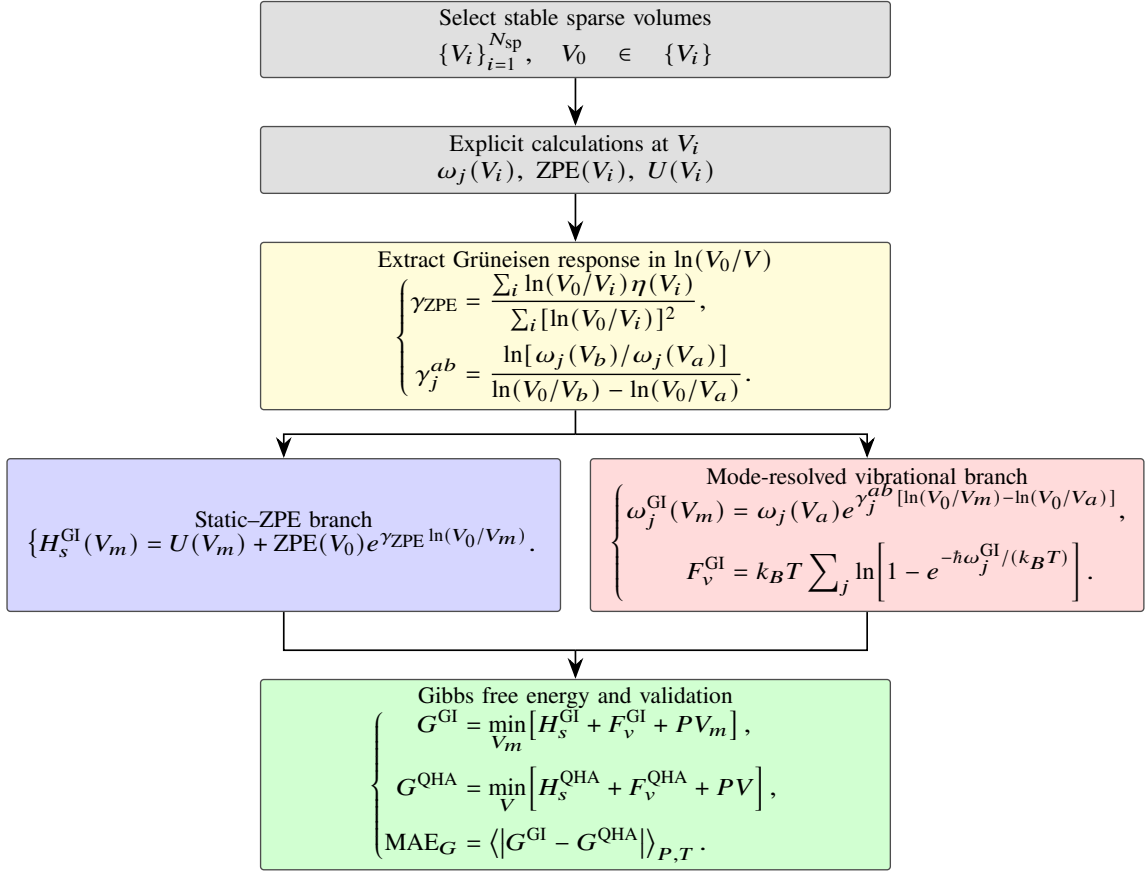
\begin{figure*}[!tbp]
\centering
\resizebox{0.84\textwidth}{!}{%
\begin{tikzpicture}[
  node distance=5mm and 3mm,
  title/.style={font=\bfseries\footnotesize, align=center},
  box/.style={
    draw=black!70,
    rounded corners=1pt,
    align=center,
    minimum height=7mm,
    inner sep=2.2pt,
    font=\footnotesize
  },
  mainbox/.style={box, minimum width=66mm, text width=62mm, font=\scriptsize},
  graybox/.style={mainbox, fill=black!12},
  yellowbox/.style={mainbox, fill=yellow!18},
  bluebox/.style={box, fill=blue!16, minimum width=58mm, text width=54mm, minimum height=16mm, font=\scriptsize},
  pinkbox/.style={box, fill=red!14, minimum width=58mm, text width=54mm, minimum height=16mm, font=\scriptsize},
  greenbox/.style={mainbox, fill=green!18},
  line/.style={line width=0.5pt},
  arrow/.style={-{Stealth[length=2.2mm]}, line width=0.5pt}
]
\node[graybox] (top)
{Select stable sparse volumes\\
$\{V_i\}_{i=1}^{N_{\mathrm{sp}}},\quad V_0\in\{V_i\}$};

\node[graybox, below=5mm of top] (sparsepoints)
{Explicit calculations at $V_i$\\
$\omega_j(V_i),\ \mathrm{ZPE}(V_i),\ U(V_i)$};

\node[yellowbox, below=5mm of sparsepoints] (response)
{Extract Gr\"uneisen response in $\ln(V_0/V)$\\[-0.2mm]
$\displaystyle
\left\{
\begin{aligned}
\gamma_{\mathrm{ZPE}}
&=\frac{
\sum_i \ln(V_0/V_i)
\eta(V_i)
}{
\sum_i[\ln(V_0/V_i)]^2
},\\
\gamma_j^{ab}
&=\frac{\ln[\omega_j(V_b)/\omega_j(V_a)]}
{\ln(V_0/V_b)-\ln(V_0/V_a)} .
\end{aligned}
\right.$};

\coordinate (branch) at ($(response.south)+(0,-5mm)$);
\node[bluebox, anchor=north east] (static) at ($(branch)+(-1.5mm,0)$)
{Static--ZPE branch\\[-0.4mm]
$
\left\{
\begin{aligned}
H_s^{\mathrm{GI}}(V_m)
&=U(V_m)+\mathrm{ZPE}(V_0)
e^{\gamma_{\mathrm{ZPE}}\ln(V_0/V_m)}.
\end{aligned}
\right.$};

\node[pinkbox, anchor=north west] (vib) at ($(branch)+(1.5mm,0)$)
{Mode-resolved vibrational branch\\[-0.4mm]
$
\left\{
\begin{aligned}
\omega_j^{\mathrm{GI}}(V_m)
&=\omega_j(V_a)e^{\gamma_j^{ab}[
\ln(V_0/V_m)-\ln(V_0/V_a)]},\\
F_v^{\mathrm{GI}}
&=k_BT\sum\nolimits_j
\ln\!\left[1-e^{-\hbar\omega_j^{\mathrm{GI}}/(k_BT)}\right].
\end{aligned}
\right.$};

\coordinate (join) at ($(response.center |- static.south)+(0,-4mm)$);
\coordinate (staticJoin) at (static.south |- join);
\coordinate (vibJoin) at (vib.south |- join);
\coordinate (split) at ($(response.south)+(0,-2.5mm)$);

\node[greenbox, anchor=north] (compare) at ($(join)+(0,-3mm)$)
{Gibbs free energy and validation\\[-0.2mm]
$\displaystyle
\left\{
\begin{aligned}
G^{\mathrm{GI}}
&=\min_{V_m}\!\left[H_s^{\mathrm{GI}}+F_v^{\mathrm{GI}}+PV_m\right],\\
G^{\mathrm{QHA}}
&=\min_V\!\left[H_s^{\mathrm{QHA}}+F_v^{\mathrm{QHA}}+PV\right],\\
\mathrm{MAE}_G
&=\left\langle\left|G^{\mathrm{GI}}-G^{\mathrm{QHA}}\right|\right\rangle_{P,T}.
\end{aligned}
\right.$};

\draw[arrow] (top.south) -- (sparsepoints.north);
\draw[arrow] (sparsepoints.south) -- (response.north);
\draw[line] (response.south) -- (split);
\draw[arrow] (split) -| (static.north);
\draw[arrow] (split) -| (vib.north);
\draw[line] (static.south) -- (staticJoin) -- (join);
\draw[line] (vib.south) -- (vibJoin) -- (join);
\draw[arrow] (join) -- (compare.north);
\end{tikzpicture}}
\caption{Schematic workflow of the reduced-volume GI
scheme. Sparse-volume phonon calculations first provide $\omega_j(V_i)$,
$\mathrm{ZPE}(V_i)$, and $U(V_i)$, with
$\eta(V_i)=\ln[\mathrm{ZPE}(V_i)/\mathrm{ZPE}(V_0)]$. The ZPE-level Gr\"uneisen parameter
$\gamma_{\mathrm{ZPE}}$ reconstructs the static--ZPE branch, while the local
mode-resolved slopes $\gamma_j^{ab}$ reconstruct the finite-temperature
vibrational branch. The superscript GI denotes the reconstructed result.
QHA on a dense volume grid is used as the benchmark reference.}
\label{fig:workflow}
\end{figure*}

Substituting Eqs.~\ref{eq:static_zpe_gi} and
\ref{eq:mode_resolved_gi} into the original Helmholtz decomposition gives
\begin{widetext}
\begin{align}
F^{\mathrm{GI}}(T,V_m)
&=
H_s^{\mathrm{GI}}(V_m)+F_v^{\mathrm{GI}}(T,V_m)
\notag\\
&=
U(V_m)
+\mathrm{ZPE}(V_0)
\exp\!\left[
\gamma_{\mathrm{ZPE}}
\ln\!\left(\frac{V_0}{V_m}\right)
\right]
\notag\\
&\quad
+k_B T\sum\nolimits_j
\ln\!\left[
1-\exp\!\left(
-\frac{\hbar\omega_j(V_a)}{k_BT}
\exp\!\left\{
\gamma_j^{ab}
\left[
\ln\!\left(\frac{V_0}{V_m}\right)
-\ln\!\left(\frac{V_0}{V_a}\right)
\right]
\right\}
\right)
\right],
\notag\\
G^{\mathrm{GI}}(P,T)
&=
\min_{V_m}\left[
F^{\mathrm{GI}}(T,V_m)+PV_m
\right].
\label{eq:free_energy_gi}
\end{align}
\end{widetext}
Here, $F^{\mathrm{GI}}$ is the reconstructed Helmholtz free energy,
$F_v^{\mathrm{GI}}$ is its finite-temperature vibrational part, and
$G^{\mathrm{GI}}$ is the Gibbs free energy obtained by minimizing over the
target volumes $V_m$ at fixed $(P,T)$.
The superscript GI labels quantities reconstructed from the sparse-volume
GI procedure and distinguishes them from QHA quantities,
which are evaluated from explicitly calculated phonon spectra on the dense
volume grid. In the three-point implementation, the sparse training volumes
are selected with respect to the equilibrium-volume path actually sampled by
the Gibbs minimization. For a target pressure-temperature window, we first
determine
\begin{equation}
V_{\mathrm{eq}}(P,T)
=
\arg\min_V\left[
F^{\mathrm{QHA}}(T,V)+PV
\right]
\end{equation}
from the dense QHA reference and identify the range traversed by
$V_{\mathrm{eq}}(P,T)$. Grid points for which the minimizing volume lies on the
lower or upper boundary of the QHA volume grid are excluded when defining this
local window and when reporting the benchmark error. Three adjacent or nearby
stable volumes are then chosen to bracket the local $V_{\mathrm{eq}}(P,T)$
path: one point on the low-volume side of the path, one point on the
high-volume side, and one point close to the center of the visited volume
interval. This local-bracketing selection keeps the GI interpolation
constrained to the volume region accessed by free-energy minimization rather
than by remote volume points far from the thermal-expansion or compression
path. For sensitivity tests, we repeat the calculation for available
three-point combinations satisfying this local bracket condition and report
the mean and standard deviation of the resulting MAE. In broad compression
benchmarks where the target window spans the full stable interval, this
procedure reduces to using representative low-, middle-, and high-volume
points. The number of sparse volumes is not a formal restriction: adding more
sparse phonon volumes gives shorter local intervals and is expected to improve
the interpolation accuracy when the additional phonon calculations are
affordable. The resulting workflow is summarized in
Fig.~\ref{fig:workflow}. In the numerical free energy evaluation, only the
retained positive-frequency modes discussed below are included in the phonon
sums.

\medskip
In the numerical free energy summations, modes with imaginary or
zero frequencies were simply discarded from the phonon contribution. For a
spectrum at volume $V$, the retained
mode set is
\begin{equation}
\mathcal{J}_+(V)
=
\left\{
j=(\vec q,n)\ \middle|\ \omega_j(V)>0
\right\}.
\end{equation}
Here, $\mathcal{J}_+(V)$ denotes the set of retained positive-frequency
phonon modes at volume $V$; modes outside this set are omitted from the
zero-point and finite-temperature vibrational sums.
The zero-point and finite-temperature vibrational terms are then evaluated as
\begin{align}
\mathrm{ZPE}(V)
&=
\frac{1}{2}
\sum_{j\in\mathcal{J}_+(V)}
\hbar\omega_j(V),
\notag\\
F_v(T,V)
&=
k_BT
\sum_{j\in\mathcal{J}_+(V)}
\ln\!\left[
1-\exp\!\left(-\frac{\hbar\omega_j(V)}{k_BT}\right)
\right].
\end{align}
For the piecewise mode-resolved GI, a local slope
$\gamma_j^{ab}$ is used only when the corresponding mode has positive
frequencies at the two endpoint volumes that define the interval.

\medskip
\noindent\textbf{Error metrics.}
The accuracy of the reduced-volume workflow was evaluated at two levels. First, the
logarithmic ZPE relation was tested from QHA ZPE data. Using the notation in
Eq.~\ref{eq:eta_zpe},
\begin{equation}
\eta_k^{\mathrm{QHA}}=\ln\!\left[\frac{\mathrm{ZPE}^{\mathrm{QHA}}(V_k)}
{\mathrm{ZPE}^{\mathrm{QHA}}(V_0)}\right],
\qquad
\eta_k^{\mathrm{GI}}=\gamma_{\mathrm{ZPE}}\ln\!\left(\frac{V_0}{V_k}\right),
\end{equation}
the ZPE-regression error is the mean absolute error (MAE)
\begin{equation}
\mathrm{MAE}_{\ln\mathrm{ZPE}}
=
\frac{1}{N_{\mathrm{val}}}\sum_{k=1}^{N_{\mathrm{val}}}
\left|\eta_k^{\mathrm{QHA}}-\eta_k^{\mathrm{GI}}\right|.
\end{equation}
Here, $\eta_k^{\mathrm{QHA}}$ and $\eta_k^{\mathrm{GI}}$ are the QHA and
GI logarithmic ZPE ratios at validation volume $V_k$,
$N_{\mathrm{val}}$ is the number of validation volume points, and
$\{V_i\}_{i=1}^{N_{\mathrm{sp}}}$ denotes the sparse volumes used to obtain
$\gamma_{\mathrm{ZPE}}$. This metric measures how well the ZPE-level effective
Gr\"uneisen slope captures the logarithmic volume dependence of the ZPE.
Second, the final thermodynamic error was quantified from the reconstructed
Gibbs free energy:
\begin{equation}
\mathrm{MAE}_{G}
=
\left\langle
\left|G^{\mathrm{GI}}(T,P)-G^{\mathrm{QHA}}(T,P)\right|
\right\rangle_{T,P}.
\end{equation}
Here, $G^{\mathrm{GI}}$ denotes the sparse-volume GI
free energy, $G^{\mathrm{QHA}}$ denotes the QHA benchmark, and
the average is taken over the sampled temperature-pressure grids.
Thus, $\mathrm{MAE}_{\ln\mathrm{ZPE}}$ evaluates the quality of the extracted
Gr\"uneisen scaling itself, whereas $\mathrm{MAE}_{G}$ evaluates the
accumulated error in the final GI free energy surface.
The pressure-temperature phase diagram is then obtained by comparing the
Gibbs free energies of all candidate phases at each sampled grid point. For a
set of phases $\mathcal{P}$, the stable phase label is assigned as
\begin{equation}
\alpha_X^*(T_k,P_k)
=
\arg\min_{\alpha\in\mathcal{P}}
G_{\alpha}^{X}(T_k,P_k),
\qquad
X\in\{\mathrm{QHA},\mathrm{GI}\}.
\label{eq:phase_argmin}
\end{equation}
Here, $\alpha$ indexes the candidate phases in the phase set $\mathcal{P}$,
$X$ specifies whether the QHA or GI free energy surface is used, and
$\alpha_X^*(T_k,P_k)$ is the stable phase label at the grid point
$(T_k,P_k)$.
Phase boundaries correspond to changes in $\alpha_X^*$ across the
P-T grids, or equivalently to near degeneracies between the
lowest competing Gibbs free energies.
For phase-diagram comparisons, the phase-label similarity is defined on the
sampled P-T grids as
\begin{equation}
\label{eq:phase_similarity}
S=
\frac{1}{N_{T,P}}
\sum_{k=1}^{N_{T,P}}
\mathcal{I}_k
\times 100\%,
\end{equation}
where $N_{T,P}$ is the number of sampled grid points and $\mathcal{I}_k=1$
when the GI and QHA stable-phase labels are identical at
$(T_k,P_k)$ and 0 otherwise.
For pointwise reconstruction maps, we use
$\Delta G_{\mathrm{GI}}=G_{\mathrm{GI}}-G_{\mathrm{QHA}}$, where
$G_{\mathrm{GI}}$ is the sparse-volume GI result.
Unless otherwise
stated, $\mathrm{MAE}_{G}$ and pointwise $\Delta G_{\mathrm{GI}}$ errors are reported in
units of meV/atom.

\subsection{Computational Details}

First-principles calculations based on density functional theory (DFT)
were performed using the Vienna \emph{Ab initio} Simulation Package
(VASP)\cite{kresse1996efficient,kresse1996efficiency}. The projector
augmented-wave (PAW) method\cite{blochl1994,kresse1999} and the
Perdew-Burke-Ernzerhof (PBE) generalized gradient
approximation\cite{perdew1996} were used.
Harmonic phonon frequencies and the related vibrational free energy were
obtained with the Phonopy package\cite{togo2015} from either
finite-displacement force constants\cite{parlinski1997} or DFPT force
constants, depending on the available dataset. When finite-displacement
calculations were used, the
displacement amplitude was 0.01~\AA{} if recorded in the Phonopy displacement
files. The QHA reference data were evaluated as
$G(V,T;P)=U(V)+F_{\mathrm{ph}}(V,T)+PV$, where $U(V)$ is the static DFT
energy and $F_{\mathrm{ph}}$ is the harmonic vibrational free energy. The
GI free energy surfaces were then generated from
sparse-volume points using the mode-resolved piecewise GI described above;
the mode Gr\"uneisen parameters were extracted
from the logarithmic volume dependence of the harmonic phonon frequencies, not
from third-order force constants.

The computational settings were system dependent. For fcc Al, a
$3\times3\times3$ supercell containing 108 atoms was used with
$\Gamma$-centered Monkhorst-Pack $k$-point meshes\cite{monkhorst1976} of
$18\times18\times18$ for the four-atom static cell and $6\times6\times6$
for the phonon calculation cell. The energy cutoff was 313~eV for the
thermal-expansion calculations and 600~eV for the compression-side
Gr\"uneisen and Gibbs free energy calculations. First-order
Methfessel-Paxton smearing\cite{methfessel1989} with a width of 0.10~eV was
used for Al. For diamond C, a $2\times2\times2$ supercell containing
64 atoms was used with an energy cutoff of 600~eV, a $\Gamma$-centered
$4\times4\times4$ DFPT $k$-point mesh, and a $16\times16\times16$ phonon
$q$ mesh.

For Si and Ge, phonon calculations used $2\times2\times2$ supercells and
$\Gamma$-centered $6\times6\times6$ force
$k$-point meshes, corresponding to an equivalent primitive-cell density of
$12\times12\times12$. The Si calculations used an energy cutoff of 520~eV,
whereas the Ge calculations used an energy cutoff of 475~eV. A
$20\times20\times20$ phonon $q$ mesh was used.

For rutile TiO$_2$ and $\beta$-PtO$_2$, the phonon calculations used
$2\times2\times2$ supercells containing 48 atoms, an energy cutoff of
600~eV, and $20\times20\times20$ phonon $q$ meshes. The force
$k$-point meshes were $4\times4\times6$ for rutile TiO$_2$ and
$4\times4\times5$ for $\beta$-PtO$_2$, with corresponding static/relaxation
meshes of $7\times7\times11$ and $7\times7\times10$, respectively. After
constructing the $2\times2\times2$ phonon supercells, the phonon-calculation
$k$-point meshes were reduced from the static/relaxation meshes by dividing
each reciprocal direction by two and rounding up when necessary.

For Ta$_2$O$_5$, VASP calculations used a 600~eV plane-wave cutoff. The
$\gamma$\cite{yang2018ta2o5}, $\gamma_1$\cite{tong2023ta2o5},
B\cite{zibrov2000ta2o5}, $\lambda$\cite{lee2013ta2o5},
L$_{\mathrm{SR}}$\cite{stephenson1971ta2o5},
$\delta$\cite{fukumoto1997ta2o5},
$\beta_{\mathrm{AL}}$\cite{aleshina2002beta},
$\beta_{\mathrm{R}}$\cite{ramprasad2003ta2o5}, and
Z\cite{zibrov2000ta2o5} polymorphs were optimized as a function of unit-cell
volume, and their Gibbs free energies were evaluated with Phonopy from the
static electronic and vibrational contributions. The structural
relaxation and static-energy $k$-point meshes were $8\times8\times2$ for
$\gamma$, $8\times8\times6$ for $\gamma_1$, $2\times4\times4$ for B,
$4\times4\times8$ for $\lambda$, $4\times2\times4$ for L$_{\mathrm{SR}}$,
$4\times4\times8$ for $\delta$ and $\beta_{\mathrm{R}}$,
$4\times8\times4$ for $\beta_{\mathrm{AL}}$, and $4\times4\times4$ for Z.
In the corresponding DFPT phonon calculations\cite{baroni2001}, the meshes were reduced by a
factor of two along each reciprocal direction: $4\times4\times1$ for
$\gamma$, $4\times4\times3$ for $\gamma_1$, $1\times2\times2$ for B,
$2\times2\times4$ for $\lambda$, $\delta$, and $\beta_{\mathrm{R}}$,
$2\times1\times2$ for L$_{\mathrm{SR}}$, $2\times4\times2$ for
$\beta_{\mathrm{AL}}$, and $2\times2\times2$ for Z. The Ta$_2$O$_5$
phonon free energies were evaluated using $16\times16\times16$ phonon
$q$ meshes for most phases.

\section{Results and Discussion}

\subsection{Representative Benchmark Systems}

\begin{figure*}[!tbp]
\centering
\includegraphics[width=0.98\textwidth]{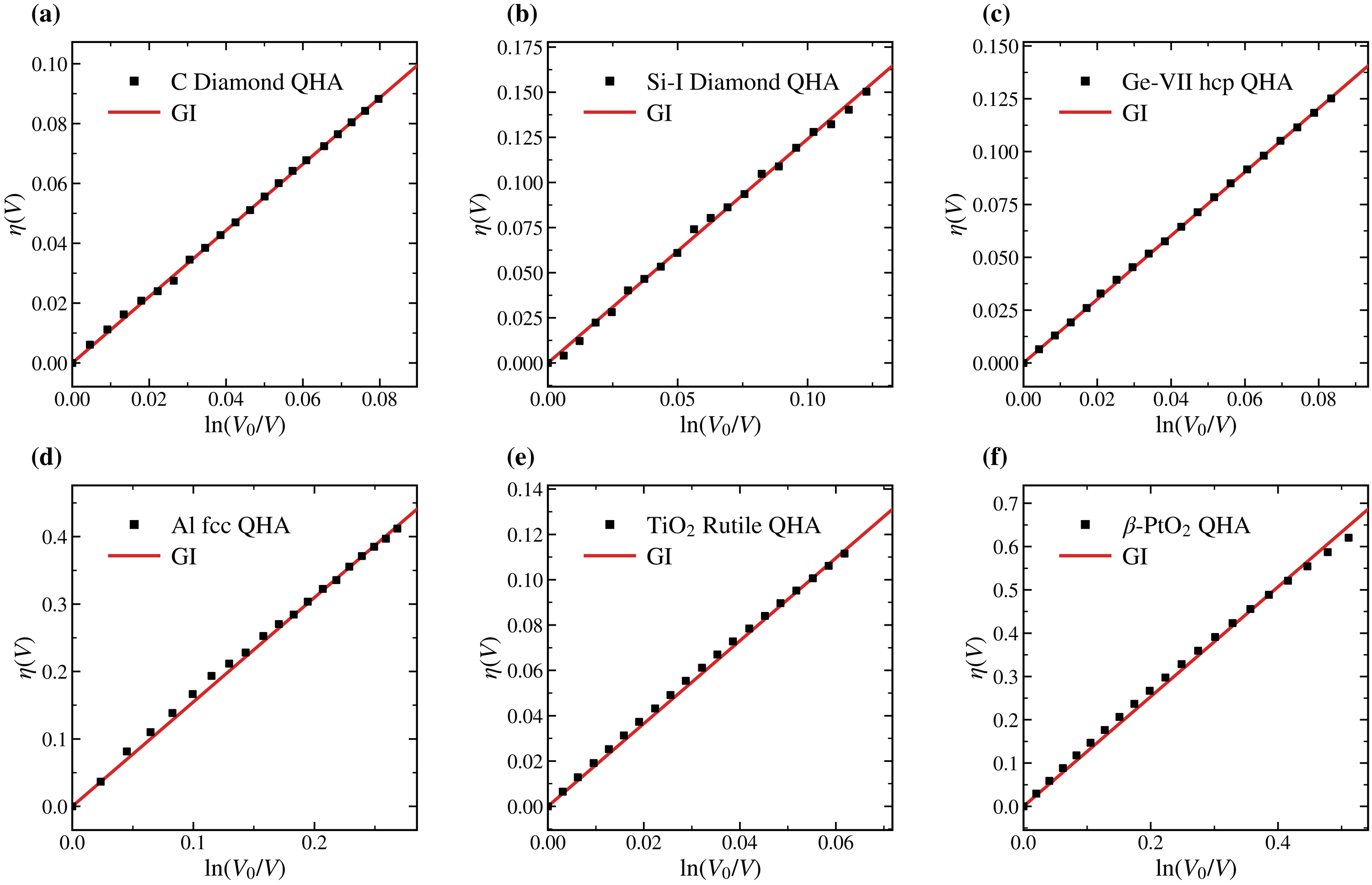}
\caption{Gr\"uneisen reconstruction for representative benchmark systems. Panels (a)--(f) show C diamond, Al fcc, Si-I diamond, Ge-VIII I4/mmm, rutile TiO$_2$, and $\beta$-PtO$_2$, respectively. In each panel, the symbols are the QHA $\eta(V)=\ln[\mathrm{ZPE}(V)/\mathrm{ZPE}(V_0)]$ values, the red line is the three-point GI regression used in this low-cost implementation, and the slope gives the ZPE-level effective Gr\"uneisen parameter used in the free energy reconstruction. For Al and Si, the compression-side GI regression uses data with $\ln(V_0/V)\ge 0$.}
\label{fig:benchmark_gamma_gi}
\end{figure*}

\begin{figure*}[!tbp]
\centering
\includegraphics[width=0.98\textwidth]{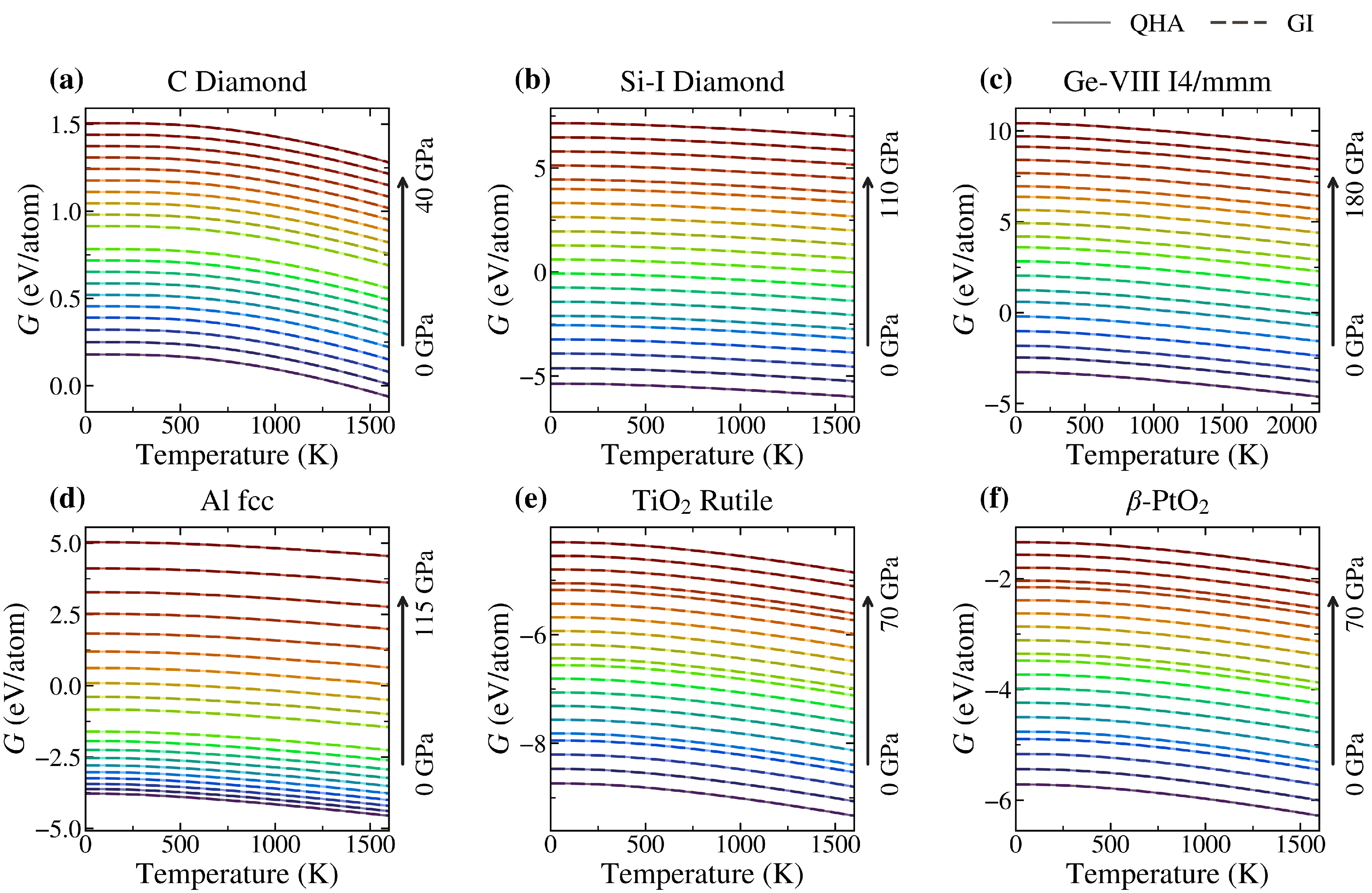}
\caption{Comparison between QHA Gibbs free energies and sparse-volume GI results for representative benchmark systems. Panels (a)--(f) correspond to C diamond, Si-I diamond, Ge-VIII I4/mmm, Al fcc, rutile TiO$_2$, and $\beta$-PtO$_2$, respectively. Solid curves denote the QHA benchmark on the dense volume grid, while dashed curves denote the present low-cost GI reconstruction. Pressure ranges are indicated outside each panel to avoid obscuring the GI curves.}
\label{fig:benchmark_gibbs_qha_gi}
\end{figure*}

\begin{figure*}[!tbp]
\centering
\includegraphics[width=0.98\textwidth]{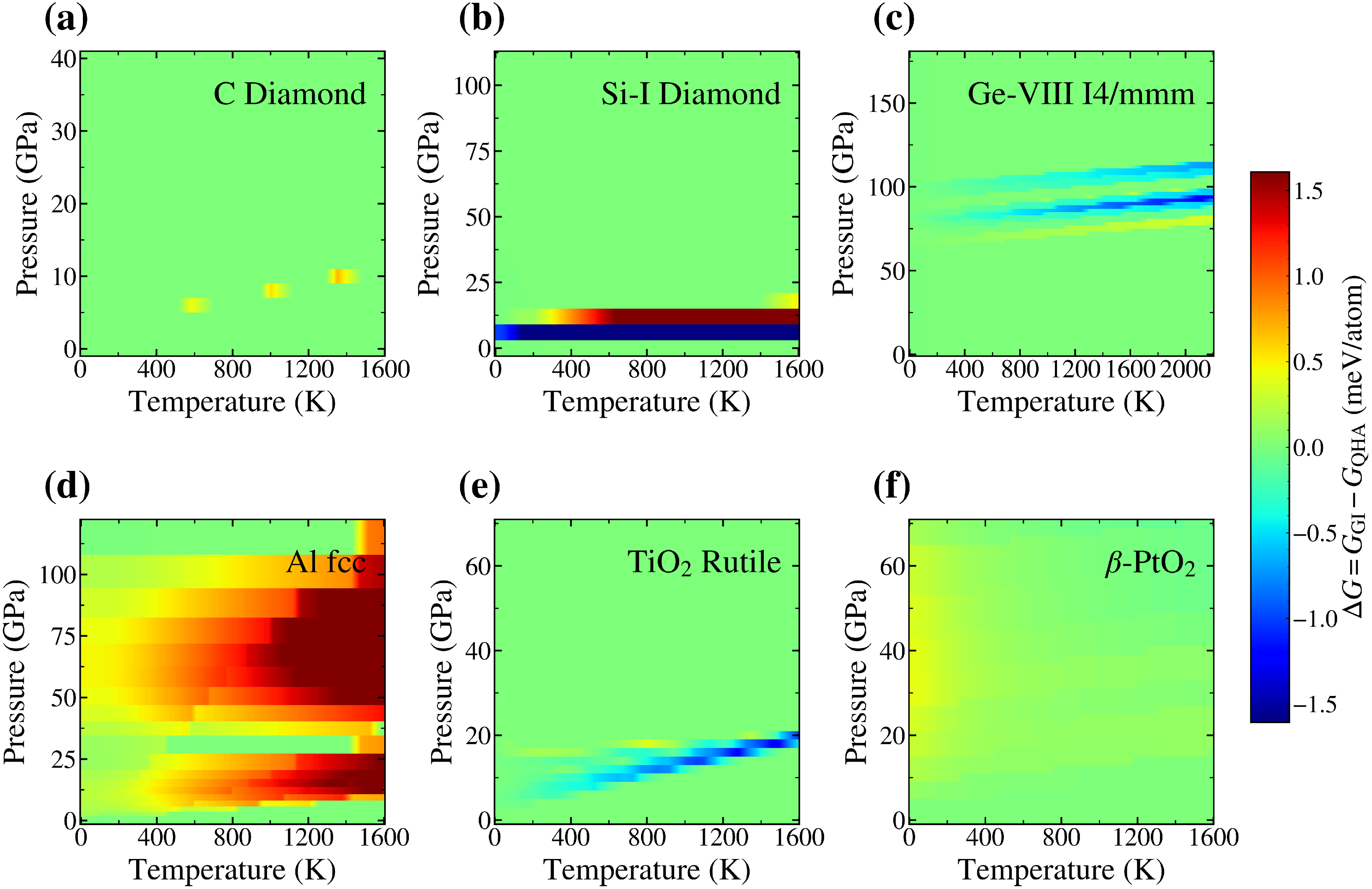}
\caption{Pointwise Gibbs free energy reconstruction-error maps for representative benchmark systems. The color scale denotes $\Delta G_{\mathrm{GI}}=G_{\mathrm{GI}}-G_{\mathrm{QHA}}$ in meV/atom over the sampled temperature-pressure grid, where $G_{\mathrm{GI}}$ is the sparse-volume GI free energy and $G_{\mathrm{QHA}}$ is the QHA benchmark. Panels (a)--(f) correspond to C diamond, Si-I diamond, Ge-VIII I4/mmm, Al fcc, rutile TiO$_2$, and $\beta$-PtO$_2$, respectively.}
\label{fig:benchmark_g_error_heatmap}
\end{figure*}

We first validate the method on typical benchmark systems with increasing
structural and chemical complexity, including diamond (C), fcc Al, Si-I
diamond, Ge-VIII I4/mmm, rutile TiO$_2$, and $\beta$-PtO$_2$.
Figure~\ref{fig:benchmark_gamma_gi} shows that
the logarithmic relation between $\eta(V)=\ln[\mathrm{ZPE}(V)/\mathrm{ZPE}(V_0)]$ and
$\ln(V_0/V)$ demonstrates good linearity for these systems, yielding the ZPE-level
Gr\"uneisen response used in the static--ZPE branch. The Gibbs free energy
comparisons in Fig.~\ref{fig:benchmark_gibbs_qha_gi} compare the reconstructed
curves with the QHA benchmark over the sampled
temperature-pressure range. The pointwise maps in
Fig.~\ref{fig:benchmark_g_error_heatmap} further show where the remaining
errors are distributed in the $(T,P)$ plane; most panels remain close to zero,
with larger deviations localized in limited pressure ranges rather than spread
uniformly over the whole grid. Quantitatively, relative to the corresponding
QHA free energies, the six representative benchmark systems have an average MAE of
0.148~meV/atom, and even the largest value remains below 0.53~meV/atom.
To keep the main comparison readable, Fig.~\ref{fig:benchmark_gibbs_qha_gi}
shows the QHA and GI curves without inset panels; representative
zoomed comparisons are provided in Appendix~\ref{app:gibbs_zoom}.

\begin{figure*}[!tbp]
\centering
\includegraphics[width=0.98\textwidth]{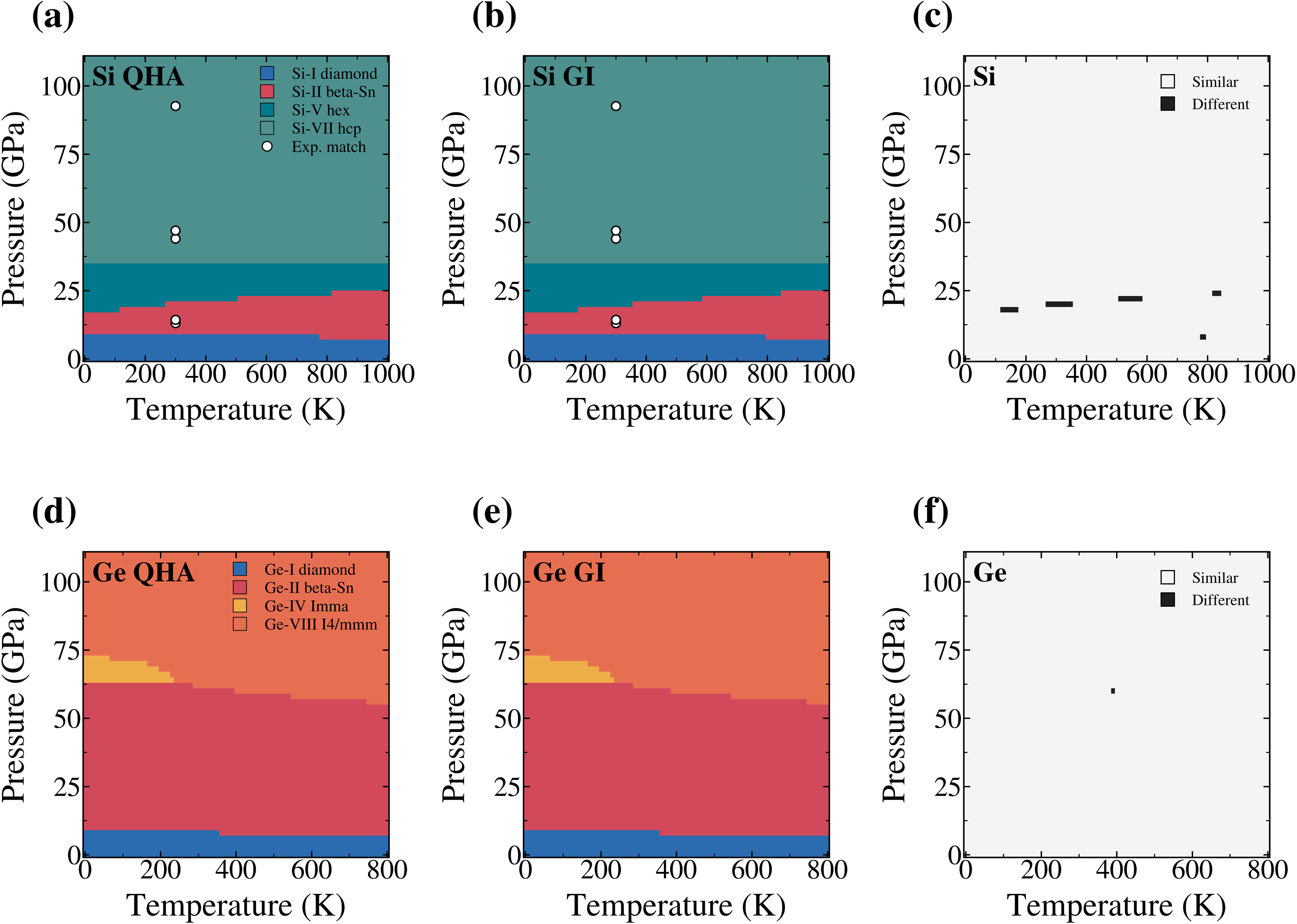}
\caption{QHA and sparse-volume GI pressure-temperature phase diagrams for Si and Ge in conservative solid-state windows. Panels (a)--(c) show the QHA Si phase diagram, GI Si phase diagram, and Si difference map, respectively; panels (d)--(f) show the corresponding QHA, GI, and difference maps for Ge. Experimental markers show only reported $(P,T)$ points whose phase assignments match the nearest-grid calculated stable phase; unmatched and melting or no-melt markers are omitted from the main figure. In the difference panels, black regions indicate grid points where the GI and QHA stable-phase labels differ; the plotted-window similarities are 99.50\% for Si and 99.98\% for Ge.}
\label{fig:si_ge_phase_diagram}
\end{figure*}

For Si and Ge, the sparse-volume GI free energy surfaces were also used to construct
P-T phase diagrams in conservative solid-state windows. As shown in
Fig.~\ref{fig:si_ge_phase_diagram}, the QHA and GI phase diagrams
preserve the main stable regions, and the remaining differences are mainly
localized near phase boundaries where small free energy deviations can
alter the assigned stable phase. The plotted windows cover 0--110~GPa and
0--1000~K for Si, and 0--110~GPa and 0--800~K for Ge, so that
liquid-related high-temperature regions are not treated as part of the
solid-state QHA comparison. The reference high-pressure phase information
for Si and Ge was taken from the studies of Anzellini \emph{et al.} and
Kelsall \emph{et al.}, respectively\cite{anzellini2019si,kelsall2021ge}.
These experimental markers are external references only and were not used
in the interpolation, free energy reconstruction, or phase-label assignment.
Open symbols mark reported $(P,T)$ points whose phase labels agree with the
nearest-grid calculated stable phase; unmatched markers and melting or
no-melt observations are omitted from the main panels. The QHA-vs-interpolated
phase-label similarities in the plotted windows are 99.50\% for Si and
99.98\% for Ge, following Eq.~\ref{eq:phase_similarity}.

\subsection{Binary-Oxide Benchmarks}

Rutile TiO$_2$ and $\beta$-PtO$_2$ extend the benchmark set from elemental and
semiconductor systems to binary oxides. Their ZPE-scaling and free energy
comparisons, included in Figs.~\ref{fig:benchmark_gamma_gi} and
\ref{fig:benchmark_gibbs_qha_gi}, show that the same low-cost sparse-volume
GI reconstruction also follows the QHA benchmark free
energies for chemically more complex oxide bonding environments. The
corresponding reconstruction errors are summarized together with the other
benchmark systems in Table~\ref{tab:reconstruction_accuracy}.

\subsection{Ta$_2$O$_5$ Polymorphs}

\begin{figure*}[!tbp]
\centering
\includegraphics[width=0.98\textwidth]{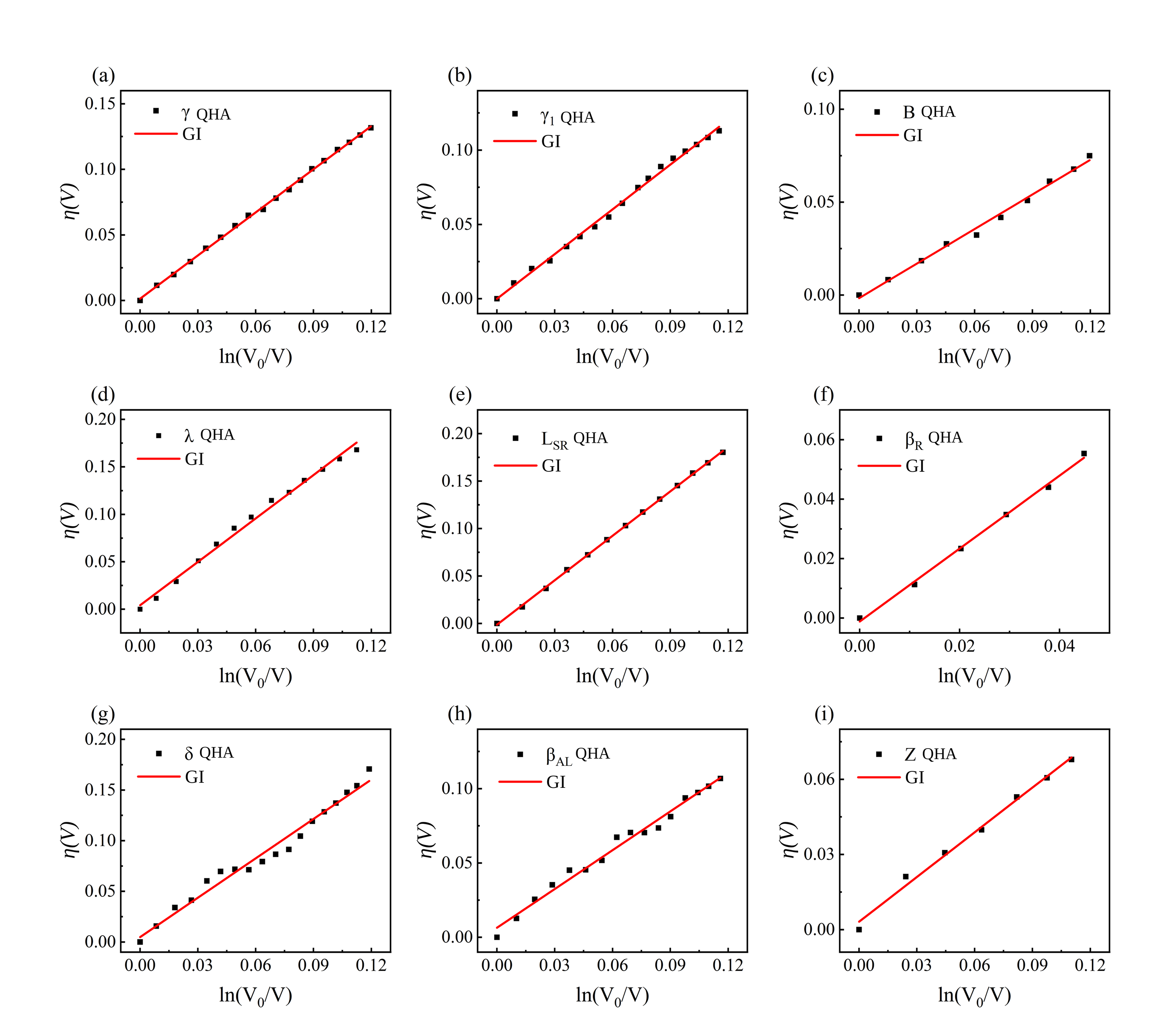}
\caption{Gr\"uneisen reconstruction for nine Ta$_2$O$_5$ polymorphs. Panels (a)--(i) correspond to $\gamma$, $\gamma_1$, B, $\lambda$, L$_{\mathrm{SR}}$, $\beta_{\mathrm{R}}$, $\delta$, $\beta_{\mathrm{AL}}$, and Z phases, respectively. In each panel, the symbols represent QHA $\eta(V)=\ln[\mathrm{ZPE}(V)/\mathrm{ZPE}(V_0)]$ data and the GI regression line gives the phase-specific effective Gr\"uneisen parameter.}
\label{fig:ta2o5_gamma_gi}
\end{figure*}

\begin{figure*}[!tbp]
\centering
\includegraphics[width=0.98\textwidth]{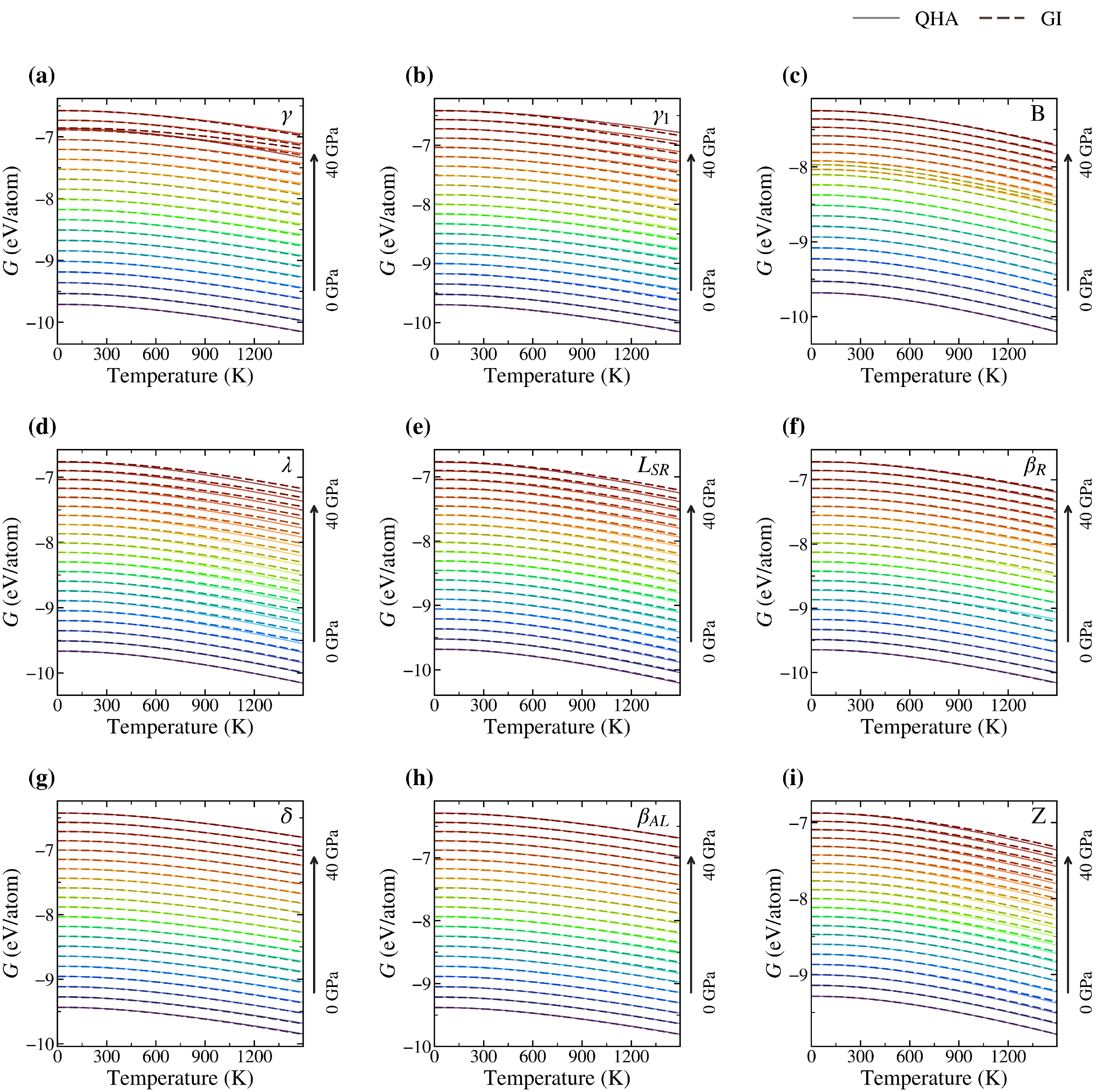}
\caption{Gibbs free energy comparison for nine Ta$_2$O$_5$ polymorphs. Panels (a)--(i) correspond to $\gamma$, $\gamma_1$, B, $\lambda$, L$_{\mathrm{SR}}$, $\beta_{\mathrm{R}}$, $\delta$, $\beta_{\mathrm{AL}}$, and Z phases, respectively. Each panel compares QHA Gibbs free energies with sparse-volume GI results under multiple pressures, with the Gibbs free energy reported in eV/atom. The GI curves capture the main QHA trends with phase-dependent errors using sparse phonon-volume points chosen by the selection procedure described in Sec.~II.}
\label{fig:ta2o5_gibbs_qha_gi}
\end{figure*}

\begin{figure*}[!tbp]
\centering
\includegraphics[width=0.98\textwidth]{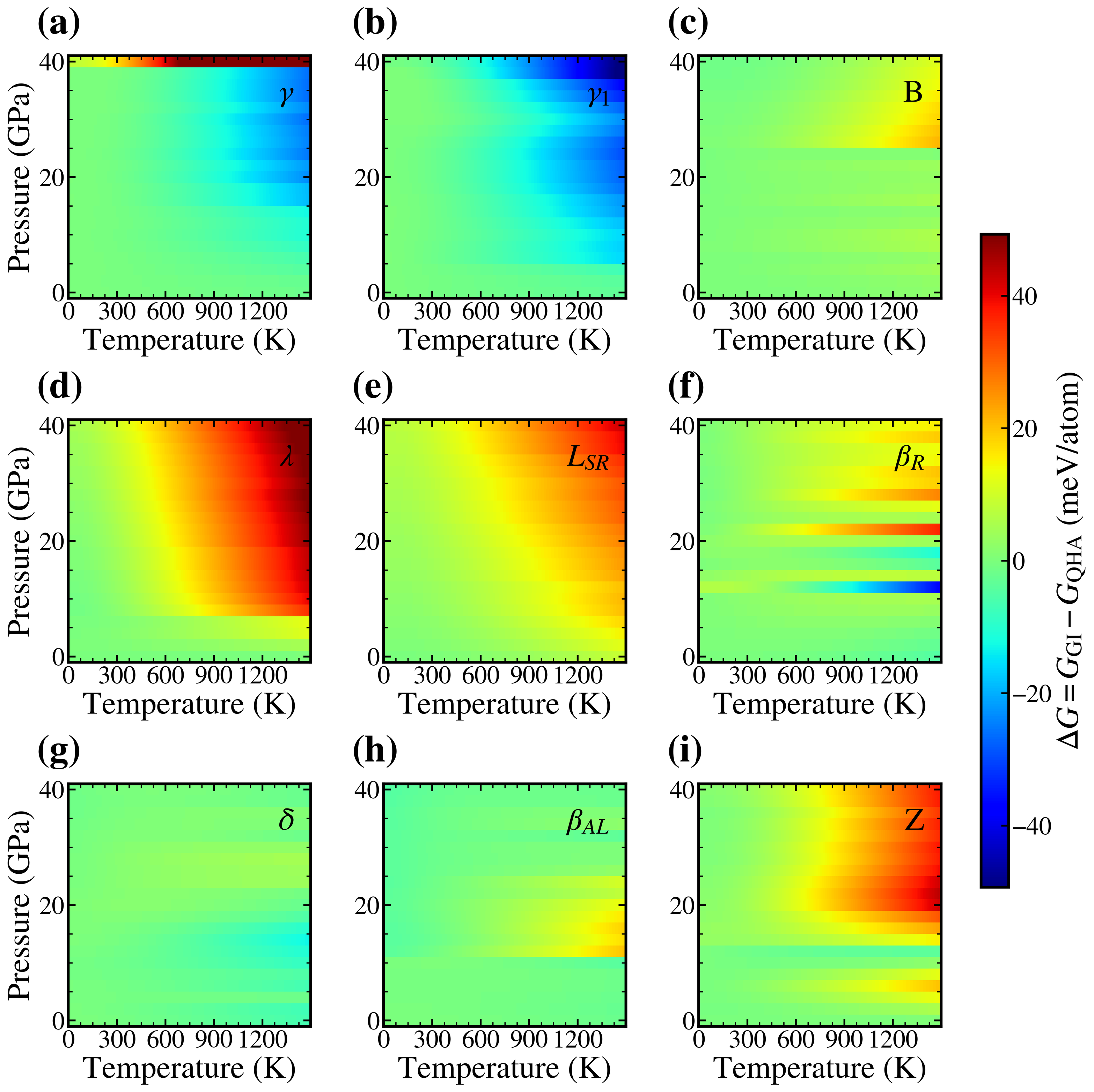}
\caption{Pointwise Gibbs free energy reconstruction-error maps for nine Ta$_2$O$_5$ polymorphs. The color scale denotes $\Delta G_{\mathrm{GI}}=G_{\mathrm{GI}}-G_{\mathrm{QHA}}$ in meV/atom over the sampled temperature-pressure grid, where $G_{\mathrm{GI}}$ is the sparse-volume GI free energy and $G_{\mathrm{QHA}}$ is the QHA benchmark. Panels (a)--(i) correspond to $\gamma$, $\gamma_1$, B, $\lambda$, L$_{\mathrm{SR}}$, $\beta_{\mathrm{R}}$, $\delta$, $\beta_{\mathrm{AL}}$, and Z phases, respectively.}
\label{fig:ta2o5_g_error_heatmap}
\end{figure*}

We next apply the same workflow to Ta$_2$O$_5$, a structurally complex
wide-gap transition metal oxide with multiple competing polymorphs. The investigated primitive cells
span $Z=1$--11 Ta$_2$O$_5$ formula units, corresponding to 7--77 atoms.
The Ta$_2$O$_5$ test is therefore a more stringent assessment of the method
than the elemental and binary benchmark systems. The nine investigated
polymorphs include B, $\beta_{\mathrm{AL}}$, $\beta_{\mathrm{R}}$, $\delta$,
$\gamma_1$, $\gamma$, L$_{\mathrm{SR}}$, $\lambda$, and Z phases. The
first-principles phase-stability data used for comparison were taken from
the ab initio phase-diagram work of Gong \emph{et al.}\cite{gongta2o5}.
Figure~\ref{fig:ta2o5_gamma_gi} shows the Gr\"uneisen reconstruction quality for
these phases, while Fig.~\ref{fig:ta2o5_gibbs_qha_gi} compares the sparse-volume
GI and QHA Gibbs free energies plotted in eV/atom. The
corresponding pointwise free energy errors are summarized in
Fig.~\ref{fig:ta2o5_g_error_heatmap}.

\begin{figure*}[!tbp]
\centering
\includegraphics[width=0.98\textwidth]{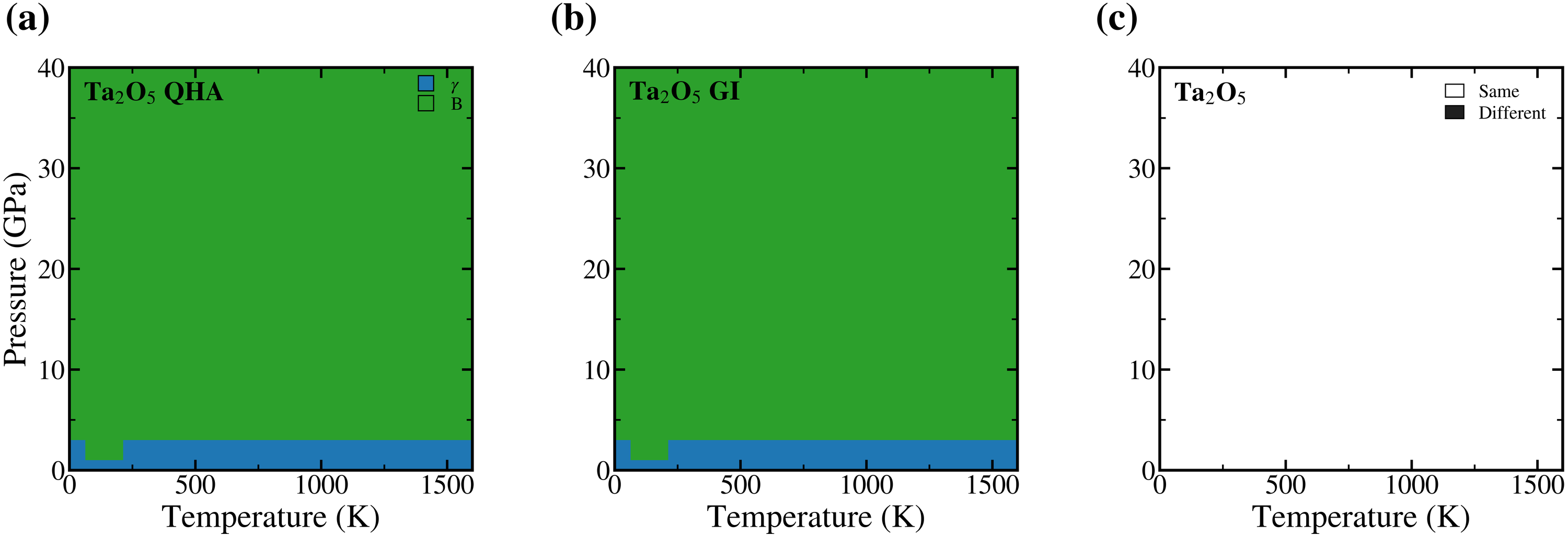}
\caption{Ta$_2$O$_5$ pressure-temperature phase diagram comparison over 0--40~GPa. Panel (a) shows the QHA phase diagram, panel (b) shows the phase diagram reconstructed from the sparse-volume GI free energy surface, and panel (c) shows the difference map between the two assignments. In this plotted range, the $\gamma$ phase is stable only in the lowest-pressure region, while the B phase dominates the higher-pressure part of the solid-state phase diagram. White and black regions in panel (c) denote identical and different GI/QHA stable-phase labels, respectively.}
\label{fig:ta2o5_phase_diagram}
\end{figure*}

The sparse-volume GI free energy surfaces also preserve the main phase-stability
topology. Figure~\ref{fig:ta2o5_phase_diagram} compares the QHA
and GI Ta$_2$O$_5$ phase diagrams over 0--40~GPa. In
this P-T window, the calculated solid-state phase diagram is
controlled by the competition between the low-pressure $\gamma$ phase and the
higher-pressure B phase. The $\gamma$ phase is confined to the lowest-pressure
region, whereas the B phase occupies most of the sampled pressure-temperature
domain. The GI phase map reproduces the same
$\gamma$--B boundary as the QHA map, and the difference panel contains
no mismatched grid points under the same grid-label definition as
Eq.~\ref{eq:phase_similarity}. The detailed reconstruction
statistics for all benchmark and Ta$_2$O$_5$ systems are summarized in
Table~\ref{tab:reconstruction_accuracy}. Across the simple benchmark systems, the
Gibbs free energy mean absolute errors relative to QHA have an average value of
0.148~meV/atom and remain below 0.53~meV/atom. For Ta$_2$O$_5$, the errors
are larger because of the more
complex polymorphic landscape, but they remain within the range needed to
reproduce the main free energy trends and phase-boundary topology. The larger
residuals are consistent with reported long-range atomic rearrangements and
lattice relaxation in Ta$_2$O$_5$-based structures, as well as the rich
high-pressure polymorphism of this oxide\cite{guo2014oxygen,yang2014longrange,gongta2o5}.

The phase-resolved Ta$_2$O$_5$ errors also provide insight into the physical
origin of the residual deviations. The error is not controlled solely by the
magnitude of the ZPE-level Gr\"uneisen parameter. For example,
$\beta_{\mathrm{R}}$ has the largest $\gamma_{\mathrm{ZPE}}$ value among the tested
Ta$_2$O$_5$ phases ($\gamma_{\mathrm{ZPE}}=2.026$) and a relatively large
$\mathrm{MAE}_{\ln\mathrm{ZPE}}$, consistent with strong volume sensitivity
of low-frequency modes. In contrast, the Z phase has a much smaller
$\gamma_{\mathrm{ZPE}}$ ($0.512$) but still shows a non-negligible error,
indicating that a small average Gr\"uneisen parameter does not guarantee a
uniformly simple phonon response. The larger errors observed for
$\lambda$, L$_{\mathrm{SR}}$, $\gamma$, and $\gamma_1$ phases are therefore better
understood as the combined effect of phase-specific soft modes, mode
crossings, and shallow free energy separations among competing polymorphs.
In such cases, the ZPE-level $\gamma_{\mathrm{ZPE}}$ captures only the average volume
response, while the piecewise mode-resolved slopes may still miss rapid
mode rearrangements if the sparse volume interval is too wide. This
interpretation explains why the GI method can still
preserve the dominant phase-stability topology while producing larger
absolute $\mathrm{MAE}_{G}$ values for selected complex polymorphs.

\begin{table*}[!tbp]
\centering
\caption{Gr\"uneisen reconstruction quality and Gibbs free energy reconstruction errors. The tabulated $\gamma_{\mathrm{ZPE}}$, coefficient of determination $R^2$, and $\mathrm{MAE}_{\ln\mathrm{ZPE}}$ refer to the ZPE-level through-origin regression, not to the full distribution of local mode-resolved slopes $\gamma_j^{ab}$. $N_{\mathrm{QHA}}$ is the number of dense QHA phonon-volume points, and $N_{\mathrm{sp}}$ is the number of sparse phonon-volume points used in GI. $\mathrm{MAE}_{G}$ measures the average absolute difference between sparse-volume GI and QHA Gibbs free energies on the temperature-pressure grid and is reported in meV/atom for all systems.}
\label{tab:reconstruction_accuracy}
\fontsize{9.3}{10.5}\selectfont
\setlength{\tabcolsep}{3.0pt}
\renewcommand{\arraystretch}{1.04}
\begin{tabular*}{\textwidth}{@{\extracolsep{\fill}}llcccccc@{}}
\toprule
\textbf{System} & \textbf{Phase} &
$\boldsymbol{\gamma}_{\mathrm{ZPE}}$ &
$\boldsymbol{R^2}$ &
\textbf{MAE$_{\ln\mathrm{ZPE}}$} &
\textbf{$N_{\mathrm{QHA}}$} &
\textbf{$N_{\mathrm{sp}}$} &
\begin{tabular}[c]{@{}c@{}}\textbf{MAE$_G$}\\ \textbf{(meV/atom)}\end{tabular} \\
\midrule
C & Diamond & 1.108182 & 0.999298 & 0.000484 & 21 & 3 & 0.004776 \\
Si & Si-I diamond & 0.713886 & 0.886645 & 0.008009 & 20 & 3 & 0.189591 \\
Al & fcc & 1.596810 & 0.980945 & 0.029276 & 21 & 3 & 0.521570 \\
Ge & Ge-VIII I4/mmm & 1.818700 & 0.998483 & 0.001942 & 20 & 3 & 0.046183 \\
TiO$_2$ & rutile & 1.826320 & 0.996902 & 0.001679 & 20 & 3 & 0.037877 \\
PtO$_2$ & $\beta$ & 1.267540 & 0.995415 & 0.011056 & 21 & 3 & 0.085837 \\
\midrule
Ta$_2$O$_5$ & B & 0.663966 & 0.990388 & 0.006216 & 21 & 3 & 3.250070 \\
Ta$_2$O$_5$ & $\beta_{\mathrm{AL}}$ & 0.955588 & 0.985694 & 0.003890 & 21 & 3 & 2.836110 \\
Ta$_2$O$_5$ & $\beta_{\mathrm{R}}$ & 2.026180 & 0.935738 & 0.021125 & 21 & 3 & 6.279970 \\
Ta$_2$O$_5$ & $\delta$ & 1.380070 & 0.979448 & 0.006873 & 21 & 3 & 2.252700 \\
Ta$_2$O$_5$ & $\gamma_1$ & 1.074820 & 0.967541 & 0.006097 & 21 & 3 & 9.481160 \\
Ta$_2$O$_5$ & $\gamma$ & 0.864189 & 0.919617 & 0.017020 & 21 & 3 & 9.150420 \\
Ta$_2$O$_5$ & L$_{\mathrm{SR}}$ & 1.503440 & 0.998527 & 0.002502 & 21 & 3 & 13.039600 \\
Ta$_2$O$_5$ & $\lambda$ & 1.299060 & 0.937292 & 0.014200 & 21 & 3 & 17.523400 \\
Ta$_2$O$_5$ & Z & 0.511854 & 0.936725 & 0.010430 & 21 & 3 & 10.937000 \\
\bottomrule
\end{tabular*}
\end{table*}

\subsection{Thermal-Expansion Coefficients of Al and Si}

\begin{figure*}[!tbp]
\centering
\includegraphics[width=\textwidth]{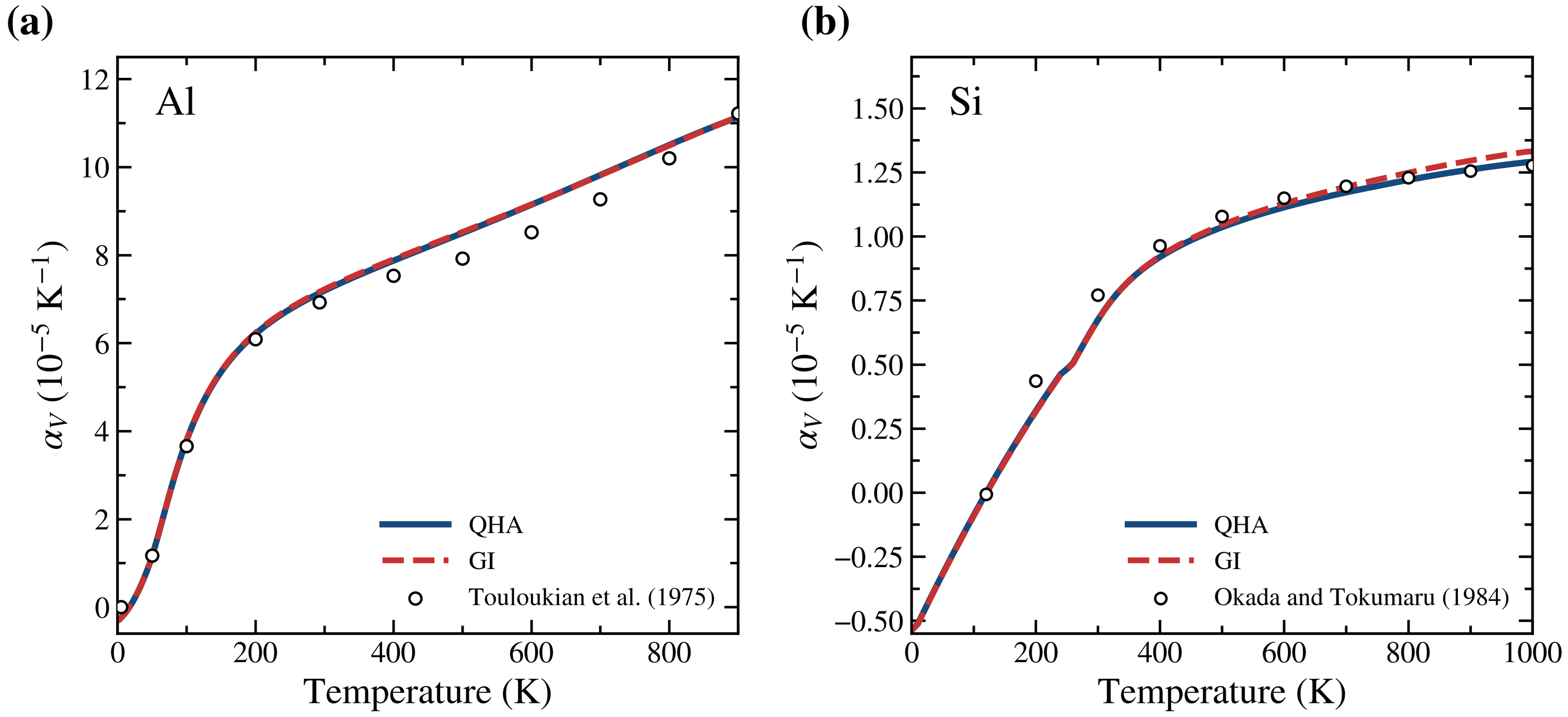}
\caption{Volumetric thermal-expansion coefficients of Al and diamond Si as functions of temperature. Panel (a) shows Al fcc and panel (b) shows Si diamond. Blue solid curves denote QHA results, while red dashed curves denote results from the sparse-volume piecewise mode-resolved GI scheme. Open circles for Al are the recommended experimental data of Touloukian \emph{et al.}\cite{touloukian1975}; experimental data for Si are obtained from the empirical linear thermal-expansion coefficient of Okada and Tokumaru\cite{okada1984} and converted using $\alpha_V=3\alpha_L$.}
\label{fig:thermal_expansion}
\end{figure*}

To further test whether the GI free energy
landscape preserves thermodynamic information, we compare the volumetric
thermal expansion coefficients of two typical systems: Al and diamond Si. The
data obtained from QHA are compared with those obtained from the piecewise
mode-resolved GI scheme. As shown in
Fig.~\ref{fig:thermal_expansion}, the experimental reference data for Al are
taken from the values compiled by Touloukian
\emph{et al.}\cite{touloukian1975}. For Si, the experimental data are taken
from the empirical thermal-expansion coefficient of high-purity silicon
reported by Okada and Tokumaru\cite{okada1984}, with the linear coefficient
converted to the volumetric coefficient by $\alpha_V=3\alpha_L$. The GI
curves closely reproduce the corresponding QHA results for both Al and
Si, including the negative-to-positive thermal-expansion trend of Si. The
discrete experimental points are presented for judging whether the calculated
thermal-expansion trend has the correct magnitude. For Al, the theoretical
curves compare well with the Touloukian data over the plotted temperature
range despite small high-temperature deviations. For Si, the comparison
mainly tests whether the method preserves the QHA negative-to-positive
crossover rather than matching each experimental point individually. This
agreement shows that the sparse mode-resolved GI can recover the QHA
thermal-expansion behavior using only a small number of phonon volume points;
the remaining differences from experiment mainly reflect the underlying
DFT/QHA accuracy rather than the interpolation procedure.

\begin{figure*}[!tbp]
\centering
\includegraphics[width=\textwidth]{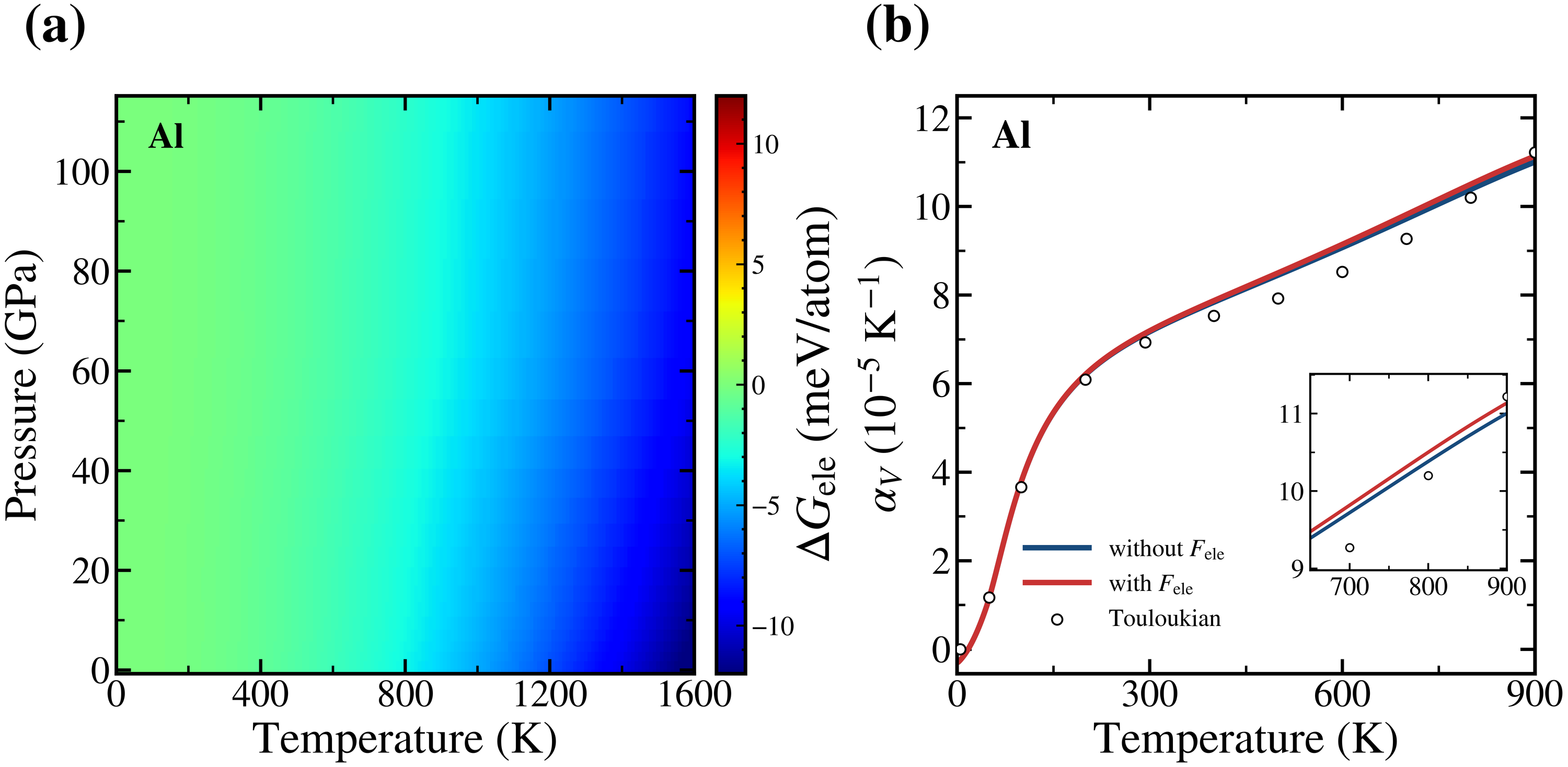}
\caption{Effect of the electronic free energy contribution in Al. Panel (a)
shows the color-map quantity
$\Delta G_{\mathrm{ele}}=G(\mathrm{with}\ \Delta F_{\mathrm{ele}})
-G(\mathrm{without}\ \Delta F_{\mathrm{ele}})$ in meV/atom over the sampled
temperature-pressure grid; negative values indicate that the electronic
contribution lowers the Gibbs free energy. Panel (b) compares the
corresponding volumetric thermal-expansion coefficient $\alpha_V(T)$ with and
without the electronic free energy correction and with the Touloukian
experimental data. The electronic term gives a small high-temperature
correction without changing the overall QHA trend.}
\label{fig:al_electronic_entropy}
\end{figure*}

For metallic Al, we also examined the electronic free energy correction
$\Delta F_{\mathrm{ele}}$ due to thermal excitation because it can slightly change the high-temperature
Gibbs free energy and the derived thermal-expansion coefficient. As shown in
Fig.~\ref{fig:al_electronic_entropy}, we define the electronic correction as
\begin{equation}
\Delta G_{\mathrm{ele}}
=G(\mathrm{with}\ \Delta F_{\mathrm{ele}})
-G(\mathrm{without}\ \Delta F_{\mathrm{ele}}).
\end{equation}
Negative
$\Delta G_{\mathrm{ele}}$ values mean that adding the electronic entropy
lowers the Gibbs free energy. The correction is modest compared with the
dominant phonon contribution, but it marginally increases $\alpha_V(T)$ at
high temperature and improves the consistency with the experimental
high-temperature trend. The corresponding DOS and entropy analysis are presented in
Appendix~\ref{app:al_electronic_entropy_dos}. In that analysis, the Al DOS is
evaluated as $D[V_{\mathrm{eq}}(T),E]$ along the thermal-expansion path, so the
volume change with temperature is included. The 0--800~K DOS curves remain very
similar, while $S_{\mathrm{ele}}$ increases nearly monotonically with
temperature. Plotting $-TS_{\mathrm{ele}}$ together with
$S_{\mathrm{ele}}$ converts the entropy into the energy scale that appears
in the electronic free energy; since $S_{\mathrm{ele}}>0$, this term
lowers the free energy at finite temperature. At 800~K,
$S_{\mathrm{ele}}=0.09645\,k_B$/atom, giving
$-TS_{\mathrm{ele}}=-6.649$~meV/atom. After the electronic internal-energy
change is included, the net electronic free energy correction entering the Al
QHA/GI free energy is about $-3.315$~meV/atom.

\subsection{CPU-Time Efficiency Analysis}

The acceleration originates from replacing dense-volume QHA
phonon calculations by phonon calculations at a reduced set of volumes used
for GI. Table~\ref{tab:cpu_time} summarizes the
central processing unit (CPU)-time comparison for the
systems investigated here. The QHA benchmark workflows require
20--21 explicit volume points, whereas the sparse-volume GI workflow uses
three sparse points in the
present low-cost implementation. Additional sparse volumes can be used when
higher accuracy is needed, at the expense of a smaller speedup. The resulting
speedups range
from 5.911$\times$ for Al to 9.023$\times$ for Si, while Ta$_2$O$_5$
still retains an 8.103$\times$ speedup despite its larger structural
complexity.

\begin{table*}[!tbp]
\centering
\caption{CPU-time comparison between the QHA benchmark and the present sparse-volume GI workflow. The point counts refer to phonon-volume calculations. CPU totals include the static $U(V)$ calculation step and the phonon jobs for this QHA/GI workflow comparison.}
\label{tab:cpu_time}
\fontsize{9.3}{10.5}\selectfont
\setlength{\tabcolsep}{3.0pt}
\renewcommand{\arraystretch}{1.04}
\begin{tabular*}{\textwidth}{@{\extracolsep{\fill}}lccccc@{}}
\toprule
\textbf{System} &
\begin{tabular}[c]{@{}c@{}}\textbf{QHA phonon}\\ \textbf{volume points}\end{tabular} &
\begin{tabular}[c]{@{}c@{}}\textbf{Sparse phonon}\\ \textbf{volume points}\end{tabular} &
\begin{tabular}[c]{@{}c@{}}\textbf{QHA}\\ \textbf{CPU time (h)}\end{tabular} &
\begin{tabular}[c]{@{}c@{}}\textbf{Sparse-GI}\\ \textbf{CPU time (h)}\end{tabular} &
\textbf{Speedup} \\
\midrule
C diamond & 21 & 3 & 1.230 & 0.176 & 7.000$\times$ \\
Al & 21 & 3 & 22.791 & 3.855 & 5.911$\times$ \\
Si & 20 & 3 & 8.628 & 0.956 & 9.023$\times$ \\
Ge & 20 & 3 & 7.305 & 1.123 & 6.504$\times$ \\
Rutile TiO$_2$ & 20 & 3 & 7.928 & 1.158 & 6.846$\times$ \\
$\beta$-PtO$_2$ & 21 & 3 & 11.131 & 1.663 & 6.695$\times$ \\
Ta$_2$O$_5$ & 20 & 3 & 40.521 & 5.001 & 8.103$\times$ \\
\bottomrule
\end{tabular*}
\end{table*}

We have further compared the present GI workflow with the VIP strategy of
Hashimoto \emph{et al.}\cite{hashimoto2025} in a low-positive-pressure
compression-side window. This benchmark skips the $P=0$ point and uses the
first three nonzero pressures for each system, avoiding boundary minimization
on the QHA volume grid over 0--800~K. In both cases the QHA boundary fraction
is zero. For Al, one- and two-random-structure VIP settings give
Gibbs free energy MAEs of 1.055 and 0.729~meV/atom, whereas two- and
three-volume GI settings give 0.0326 and 0.00349~meV/atom. For Si, the
corresponding VIP MAEs are 0.254 and 0.183~meV/atom, whereas two- and
three-volume GI settings give 0.0144 and 0.0103~meV/atom. The detailed
compression-side table, together with the separate $P=0$ thermal-expansion
comparison that is most favorable to VIP, is provided in
Appendix~\ref{app:vip_gi_comparison}.

\subsection{Applicable Range of the Method}

\begin{figure*}[!tbp]
\centering
\includegraphics[width=0.64\textwidth]{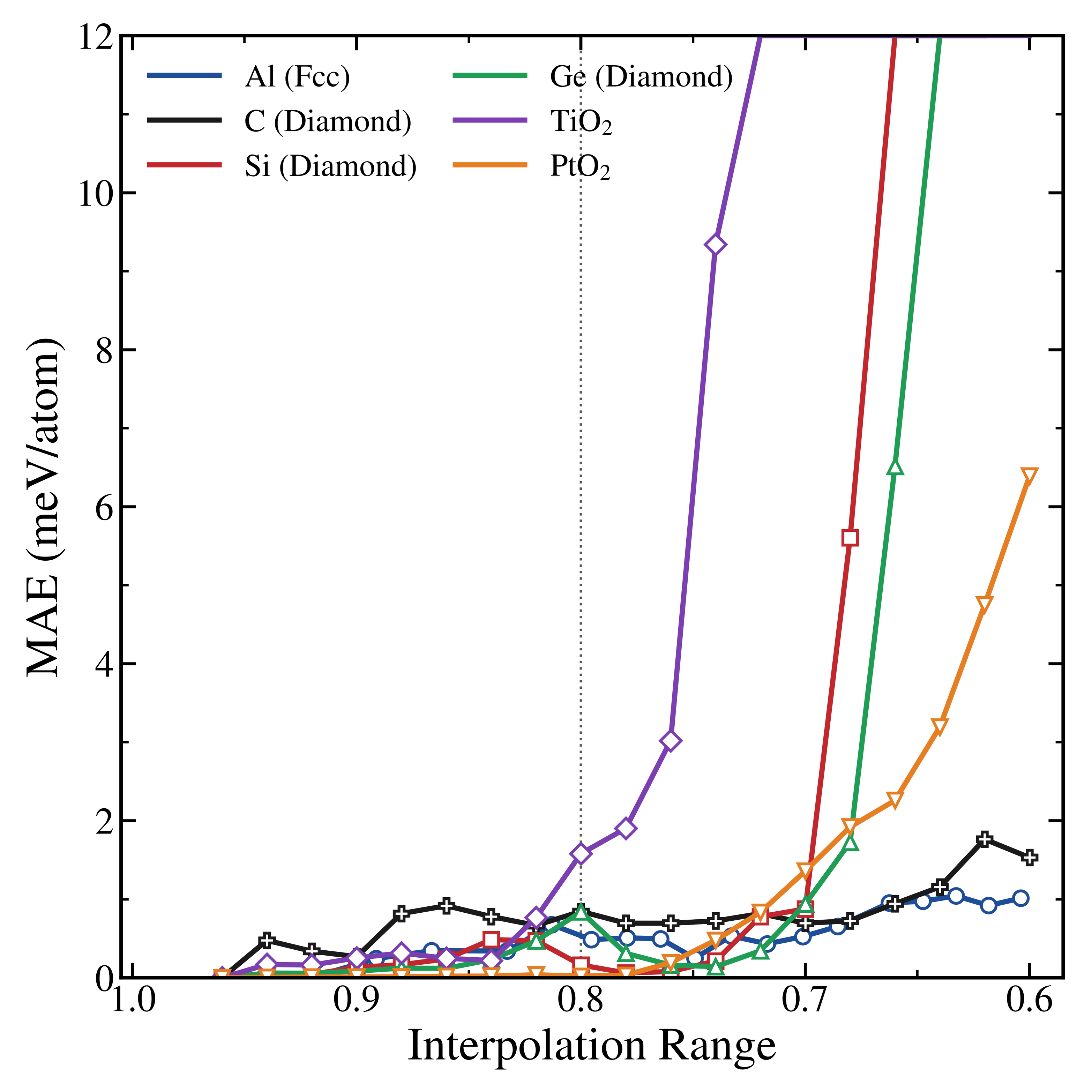}
\caption{Dependence of Gibbs free energy MAE on GI range. The curves show Al fcc, C diamond, Si diamond, Ge diamond, rutile TiO$_2$, and $\beta$-PtO$_2$. The horizontal axis denotes the lower volume bound $V_{\min}/V_0$ of the reconstructed range; moving to the right corresponds to extending the reconstruction to stronger compression. The rapid error increase at strong compression indicates the practical applicability limit of the sparse-volume GI assumption.}
\label{fig:interpolation_range}
\end{figure*}

The reliability of the reduced-volume reconstruction depends on the range of
volume compression or expansion within which the phonon spectra and extracted
Gr\"uneisen responses vary smoothly. To quantify this range, we evaluate the
Gibbs free energy MAE as GI is gradually extended from volumes close to $V_0$ toward
approximately $0.8V_0$ and below. Figure~\ref{fig:interpolation_range}
shows that most systems maintain small errors down to about
$0.8V_0$, whereas the error increases rapidly once the reconstruction range extends
to stronger compression. This behavior defines the practical range of the
method: it is reliable for moderate compression where the dominant
volume effect remains smooth enough for sparse-volume GI, and caution is
required when extrapolating to substantially larger compressions. Physically,
the rapid error growth beyond this range indicates the breakdown of the smooth
GI assumption. At
strong compression, higher-order volume terms of the interatomic potentials,
mode-dependent Gr\"uneisen parameters, mode crossings, and incipient structural transformations can
change the topology of the phonon density of
states\cite{allen2020,masuki2022}. Once this occurs, the compressed spectrum
can no longer be reconstructed reliably by sparse local Gr\"uneisen slopes or
by a single scalar scaling factor, and the method should be treated as a
validated GI tool rather than as an uncontrolled
extrapolation scheme.

\section{Conclusion}

In this work, we developed a reduced-volume GI
strategy for accelerating phonon-related Gibbs free energy calculations under
volume compression. Phonon spectra are calculated only at a few stable volumes
and are then used to extract the ZPE-level Gr\"uneisen parameter and local
mode-resolved Gr\"uneisen slopes. These extracted parameters reconstruct the
static--ZPE branch, the finite-temperature phonon branch, and the Gibbs
free energy surface on the studied volume grids. This efficiently reduces the
computational cost of Gibbs free energy and P-T phase diagram
construction while retaining the dominant volume dependence of the phonon
contribution.

Applications on benchmark systems ranging from simple metals and
semiconductors to much more complex transition-metal oxides such as
Ta$_2$O$_5$ polymorphs demonstrate the reliability of the method for
nonmagnetic crystalline materials.
For diamond, Al, Si, Ge, rutile TiO$_2$, and $\beta$-PtO$_2$, the sparse-volume
GI Gibbs free energies agree well with the QHA benchmark, with a six-system
average $\mathrm{MAE}_G$ of 0.148~meV/atom and a maximum value of
0.522~meV/atom. For
Ta$_2$O$_5$, the method
captures the main free energy variation trends and phase-stability topology
across nine polymorphs. The same GI free energy surfaces also provide a good
description of thermal expansion coefficients for Al and Si. In terms of
efficiency, the present implementation reduces the number of explicit phonon
calculation points from about 20--21 to 3 and achieves
speedups of 5.911--9.023$\times$ in the tested systems, with Ta$_2$O$_5$
still showing an 8.103$\times$ reduction in cost. This three-point
choice is not a methodological restriction; increasing the number of sparse-volume
points provides a straightforward route to improve accuracy when additional phonon
calculations are acceptable.

The method is most suitable for moderate compression ranges where the
ZPE-level and mode-resolved Gr\"uneisen responses remain smooth and where the
vibrational contribution is the dominant finite-temperature term. Systems in
which magnetic entropy, strong electronic-correlation effects, or large
configurational entropy make comparable contributions require additional
free energy terms and are outside the scope of the present work.
Within the specified range of volume variation, the method provides a practical route for rapid
high-temperature and high-pressure phase-diagram construction in materials where
conventional volume-dependent QHA calculations are computationally
demanding.

\appendix
\setcounter{figure}{0}
\setcounter{table}{0}
\renewcommand{\thefigure}{A\arabic{figure}}
\renewcommand{\thetable}{A\Roman{table}}
\section{Zoomed Gibbs Free Energy Comparisons}
\label{app:gibbs_zoom}

\begin{figure*}[!tbp]
\centering
\includegraphics[width=0.96\textwidth]{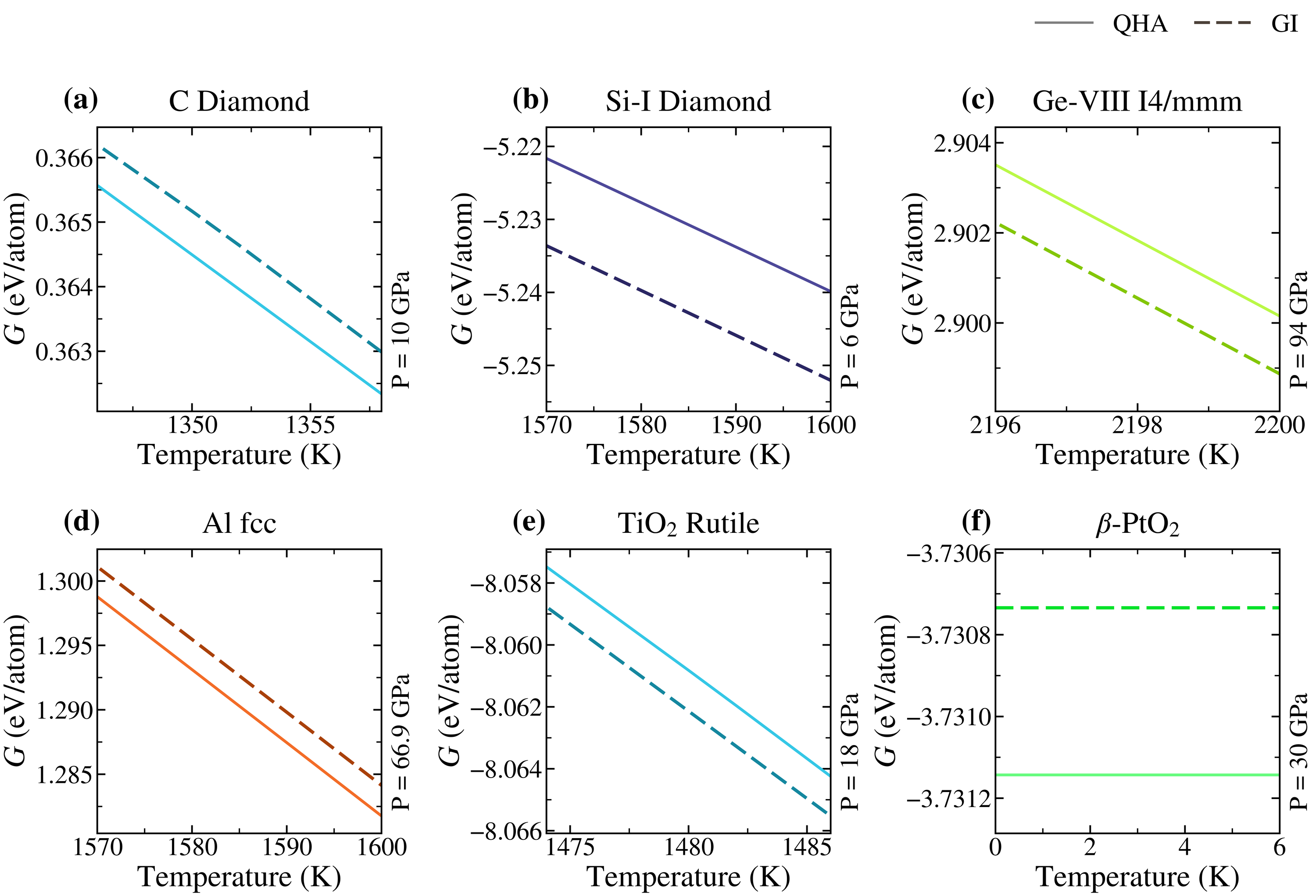}
\caption{Zoomed Gibbs free energy comparisons corresponding to the benchmark
systems in Fig.~\ref{fig:benchmark_gibbs_qha_gi}. Solid curves denote QHA
results and dashed curves denote sparse-volume GI results.}
\label{fig:gibbs_zoom_appendix}
\end{figure*}

Figure~\ref{fig:gibbs_zoom_appendix} gives zoomed comparisons of selected
QHA and GI Gibbs free energy curves from
Fig.~\ref{fig:benchmark_gibbs_qha_gi}. These panels are placed in the appendix
so that the main figure keeps the full temperature range and avoids overlap
between inset axes, curve labels, and plotted data.

\section{VIP--GI Compression and Expansion Comparisons}
\label{app:vip_gi_comparison}

\begin{figure*}[!tbp]
\centering
\includegraphics[width=0.88\textwidth]{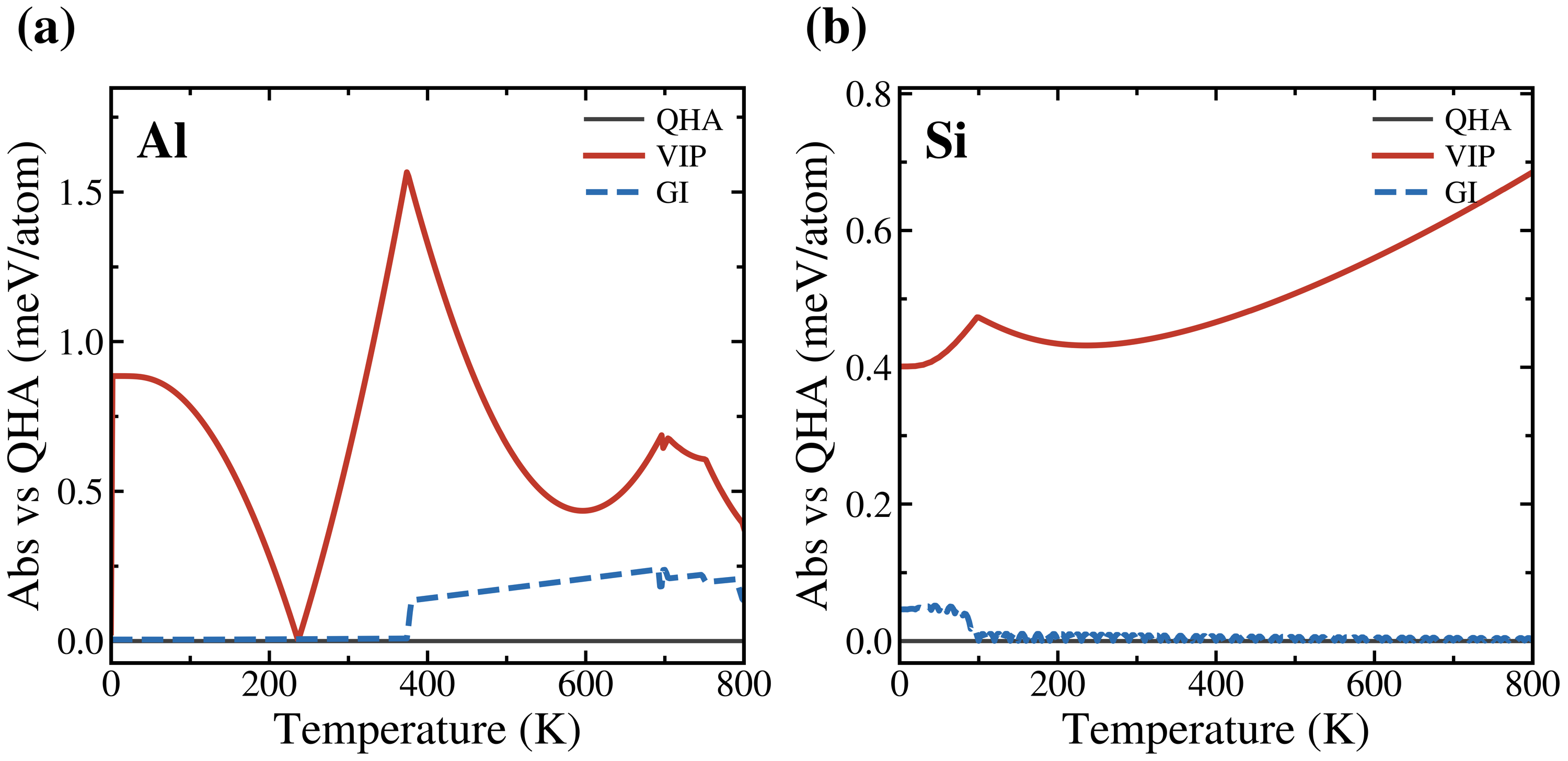}
\caption{Compression-side absolute Gibbs free energy error relative to QHA for
VIP and GI in the low-positive-pressure window. Panels (a) and (b) show Al and
Si, respectively, using the pressure windows described in the text. This
comparison excludes the $P=0$ boundary region and tests the methods on the
low-pressure compressed side.}
\label{fig:vip_gi_error_compression}
\end{figure*}

\begin{figure*}[!tbp]
\centering
\includegraphics[width=0.88\textwidth]{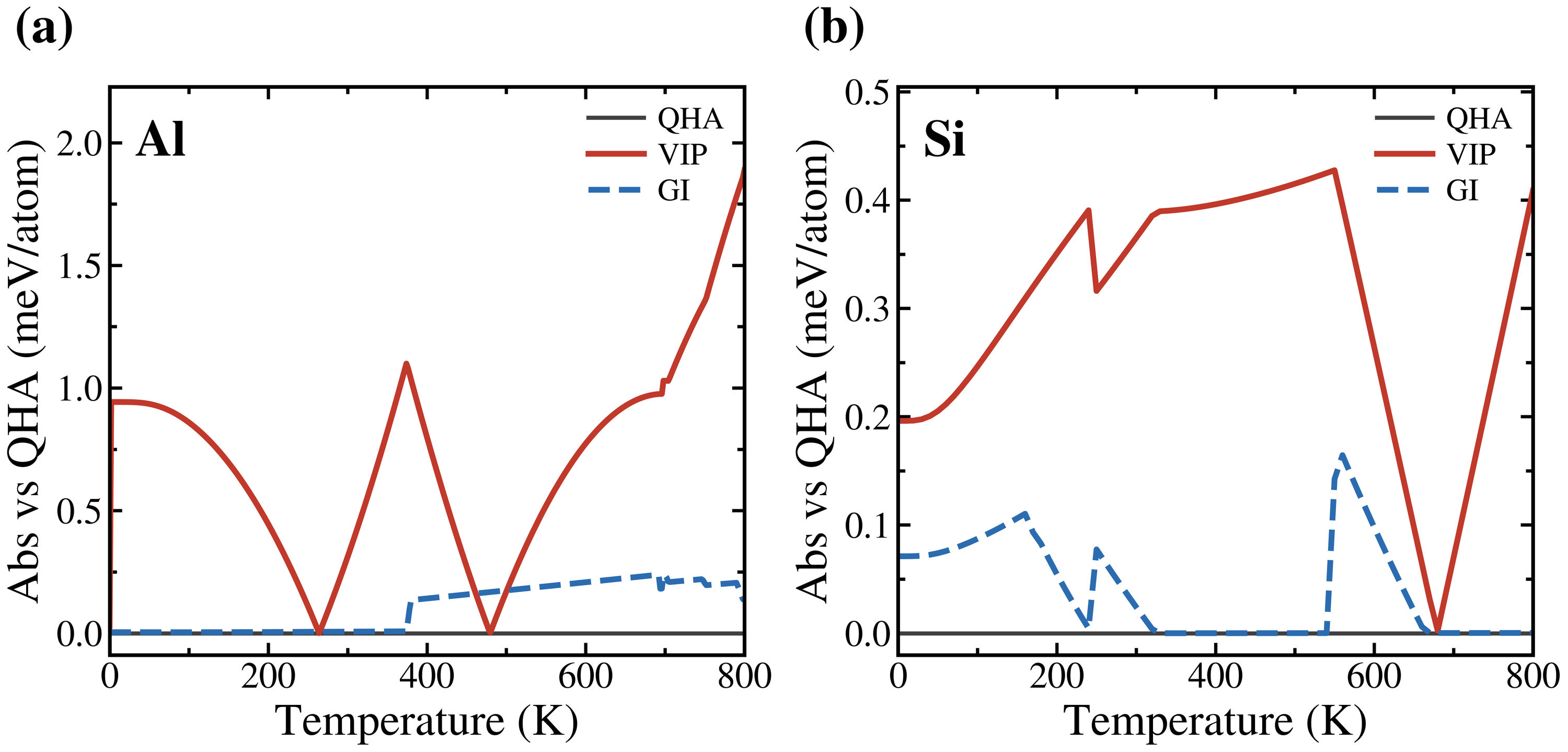}
\caption{Expansion-side absolute Gibbs free energy error relative to QHA for
VIP and GI along the $P=0$~GPa thermal-expansion path. Panels (a) and (b) show
Al and Si, respectively. Under this VIP-favorable setting, the GI curves use
local sparse volumes selected along the same thermal-expansion path and remain
closer to the QHA reference over most of the tested temperature range.}
\label{fig:vip_gi_error_expansion}
\end{figure*}

The VIP--GI comparison is reported in two steps. We first test a
low-positive-pressure compression-side window, where the $P=0$ point is
skipped so that the QHA minimum does not sit on the high-volume boundary.
After showing that GI gives lower errors than VIP in this boundary-free
compression-side test, we also evaluate the $P=0$~GPa thermal-expansion path
emphasized in the VIP work. This second setting is included for fairness
because it is the regime where VIP is expected to be most favorable.
Figure~\ref{fig:vip_gi_error_compression} shows the compression-side
comparison, and Fig.~\ref{fig:vip_gi_error_expansion} shows the
thermal-expansion comparison.

For both settings, the reported MAE is
$\langle |G_{\mathrm{GI}}(T)-G_{\mathrm{QHA}}(T)| \rangle$ over
0--800~K in meV/atom. CPU hours were obtained from the elapsed VASP time
multiplied by the number of message passing interface (MPI) cores. These
VIP--GI tables use an incremental phonon-related cost definition: static
equation-of-state (EOS) and volume-grid static-energy calculations common to
the compared workflows are not included. This differs from
Table~\ref{tab:cpu_time}, which reports the QHA/GI workflow cost including
static $U(V)$ and phonon jobs. In the tables, ``rand.'', ``vol.'', and
``ph.'' denote random structure, selected volume point, and harmonic phonon
calculation, respectively.

The compression-side benchmark uses the first three nonzero pressure points
from the supported low-pressure grids. This removes the $P=0$ boundary issue
and gives a QHA boundary fraction of zero for both Al and Si over
$T=0$--800~K. The GI rows use sparse volumes selected within the same
low-positive-pressure volume interval, and the table reports only the number
of selected volumes. As summarized in Table~\ref{tab:vip_gi_cpu_positive_low_3p},
GI has lower MAEs than VIP for both systems in this boundary-free
compression-side window.

\begin{table*}[!tbp]
\centering
\caption{Low-positive-pressure compression-side VIP--GI MAE and CPU-time comparison for Al and Si.}
\label{tab:vip_gi_cpu_positive_low_3p}
\fontsize{9.3}{10.5}\selectfont
\setlength{\tabcolsep}{3.0pt}
\renewcommand{\arraystretch}{1.04}
\begin{tabular*}{\textwidth}{@{\extracolsep{\fill}}lllcc@{}}
\toprule
\textbf{System} & \textbf{Method} & \textbf{Input} &
\begin{tabular}[c]{@{}c@{}}\textbf{MAE vs QHA}\\ \textbf{(meV/atom)}\end{tabular} &
\begin{tabular}[c]{@{}c@{}}\textbf{CPU time}\\ \textbf{(h)}\end{tabular} \\
\midrule
Al & VIP & 1 rand. + vol. + ph. & 1.054713 & 28.829 \\
Al & VIP & 2 rand. + vol. + ph. & 0.728708 & 47.465 \\
Al & GI 2-point & 2 vol. & 0.032631 & 13.012 \\
Al & GI 3-point & 3 vol. & 0.003489 & 20.318 \\
\midrule
Si & VIP & 1 rand. + vol. + ph. & 0.254008 & 3.522 \\
Si & VIP & 2 rand. + vol. + ph. & 0.182846 & 4.601 \\
Si & GI 2-point & 2 vol. & 0.014391 & 2.826 \\
Si & GI 3-point & 3 vol. & 0.010253 & 3.948 \\
\bottomrule
\end{tabular*}
\end{table*}

Although the compression-side benchmark already shows a clear GI advantage,
VIP was originally formulated and validated mainly for thermal expansion
around a reference volume. For a fair comparison in that VIP-favorable regime,
we therefore also test the $P=0$~GPa path over 0--800~K. The complete QHA
reference was first used to determine the volume range traversed by
$V_{\mathrm{eq}}(0,T)$. The GI training volumes were then selected only within
this local thermal-expansion interval: one near the low-temperature volume,
one near the high-temperature volume, and one near the middle of the interval.
Thus, in this VIP-oriented benchmark, GI does not use remote compressed
volumes. For Si, the original 20-volume QHA grid reached the high-volume
boundary along the $P=0$ path at high temperature; five additional expanded
volumes were therefore added to construct a boundary-free QHA reference for
both VIP and GI.

\begin{table*}[!tbp]
\centering
\caption{Expansion-side VIP--GI MAE and CPU-time comparison for Al and Si along the $P=0$~GPa thermal-expansion path.}
\label{tab:vip_gi_cpu_expansion}
\fontsize{9.3}{10.5}\selectfont
\setlength{\tabcolsep}{3.0pt}
\renewcommand{\arraystretch}{1.04}
\begin{tabular*}{\textwidth}{@{\extracolsep{\fill}}lllcc@{}}
\toprule
\textbf{System} & \textbf{Method} & \textbf{Input} &
\begin{tabular}[c]{@{}c@{}}\textbf{MAE vs QHA}\\ \textbf{(meV/atom)}\end{tabular} &
\begin{tabular}[c]{@{}c@{}}\textbf{CPU time}\\ \textbf{(h)}\end{tabular} \\
\midrule
Al & VIP & 1 rand. + vol. + ph. & 1.072530 & 28.829 \\
Al & VIP & 2 rand. + vol. + ph. & 0.712393 & 47.465 \\
Al & GI 2-point & 2 vol. & 0.171588 & 13.012 \\
Al & GI 3-point & 3 vol. & 0.103539 & 20.318 \\
\midrule
Si & VIP & 1 rand. + vol. + ph. & 0.299018 & 3.522 \\
Si & VIP & 2 rand. + vol. + ph. & 0.393091 & 4.601 \\
Si & GI 2-point & 2 vol. & 0.072145 & 2.826 \\
Si & GI 3-point & 3 vol. & 0.039683 & 3.948 \\
\bottomrule
\end{tabular*}
\end{table*}

\begin{figure*}[!tbp]
\centering
\includegraphics[width=0.88\textwidth]{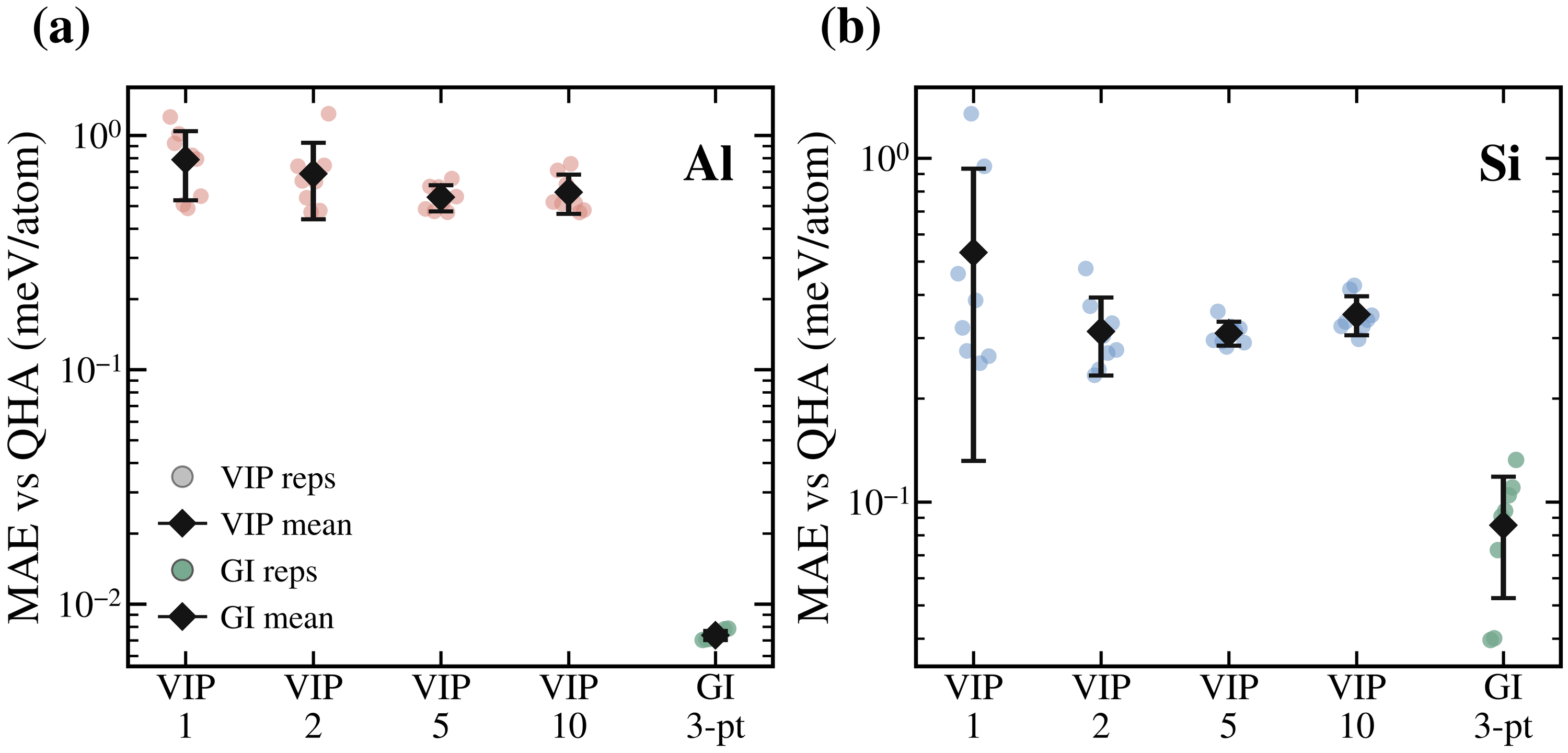}
\caption{Selection sensitivity in the expansion-side $P=0$~GPa VIP--GI
comparison. Panels (a) and (b) show Al and Si, respectively. VIP points denote
different random-subset choices, while GI points denote local three-volume
combinations satisfying the bracket condition around the
$V_{\mathrm{eq}}(0,T)$ path. Black diamonds and error bars show the mean and
standard deviation for each group.}
\label{fig:vip_gi_sensitivity}
\end{figure*}

\section{Experimental Reference Points for Thermal Expansion}
\label{app:thermal_expansion_exp}

The discrete experimental points in Fig.~\ref{fig:thermal_expansion} are used
only as external reference data and are not included in the Gr\"uneisen
interpolation reconstruction. For Al, the points are the recommended volumetric thermal
expansion data compiled by Touloukian \emph{et al.}\cite{touloukian1975}. For
Si, the experimental reference is the linear thermal-expansion coefficient of
high-purity silicon reported by Okada and Tokumaru\cite{okada1984}, converted
to the volumetric coefficient using $\alpha_V=3\alpha_L$. Therefore, the
comparison tests whether the QHA and sparse-volume GI calculations give
the correct magnitude and temperature trend. The remaining deviations from
experiment reflect the underlying DFT/QHA approximation and, for Al, the
high-temperature electronic free energy correction discussed in
Fig.~\ref{fig:al_electronic_entropy}.

\section{Electronic Entropy and DOS Analysis for Al}
\label{app:al_electronic_entropy_dos}

\begin{figure*}[!tbp]
\centering
\includegraphics[width=0.92\textwidth]{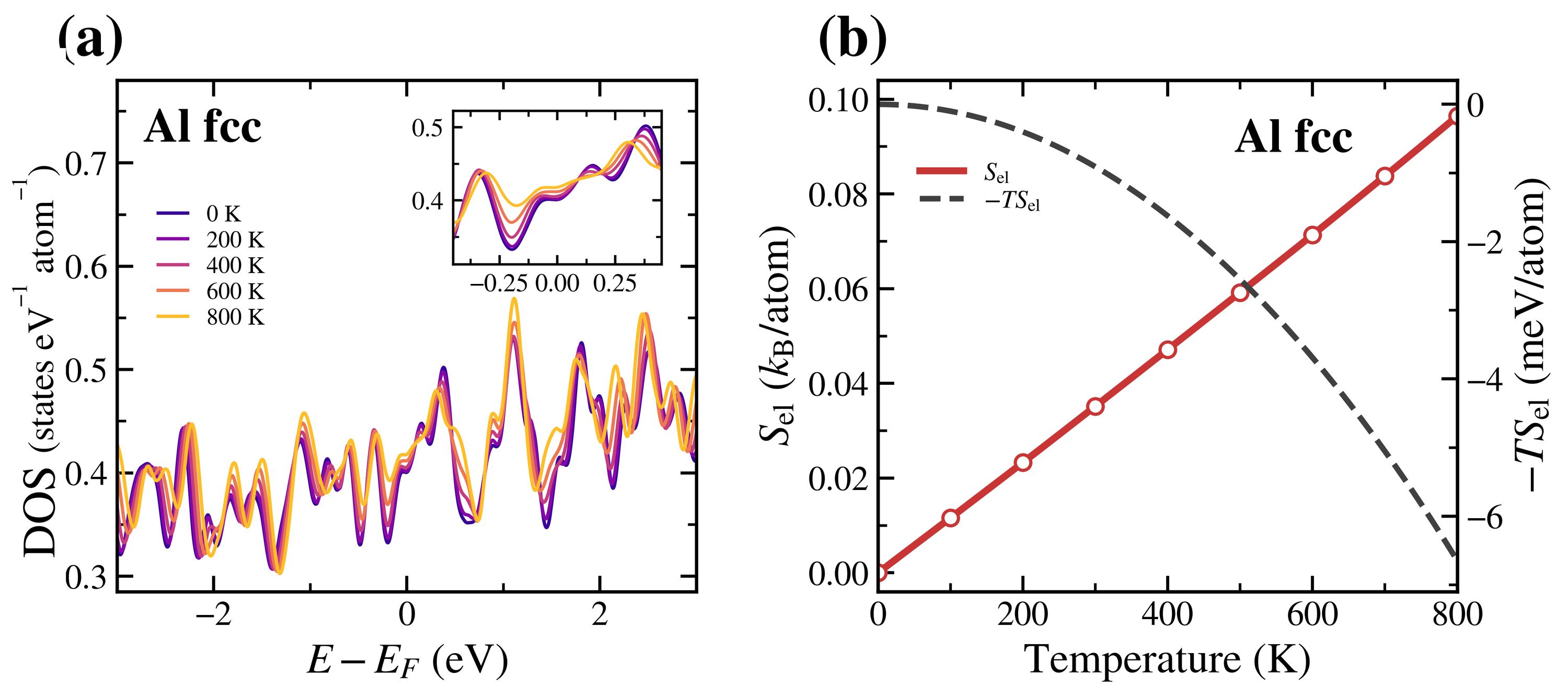}
\caption{Electronic density of states and electronic entropy of fcc Al along
the thermal-expansion path. Panel (a) shows the electronic DOS
$D[V_{\mathrm{eq}}(T),E]$, whose vertical axis is given in
units of states eV$^{-1}$ atom$^{-1}$, from 0 to 800~K, with
the energy axis aligned to the Fermi level. Panel (b) shows the electronic
entropy $S_{\mathrm{ele}}(T)$ and the corresponding free energy term
$-TS_{\mathrm{ele}}$. The latter is plotted to show the energy scale by which
electronic entropy lowers the free energy. The DOS includes the
thermal-expansion effect through the temperature-dependent equilibrium volume,
whereas explicit electron-phonon DOS broadening from thermally displaced atomic
configurations is not included.}
\label{fig:al_dos_sel_appendix}
\end{figure*}

Figure~\ref{fig:al_dos_sel_appendix} summarizes the DOS-based electronic
entropy analysis used for the Al electronic correction in
Fig.~\ref{fig:al_electronic_entropy}. The DOS curves were obtained from
ground-state static electronic DOS calculations at different volumes and then
interpolated to the equilibrium volume $V_{\mathrm{eq}}(T)$ at each
temperature. Therefore, the plotted $D[V_{\mathrm{eq}}(T),E]$ includes the
thermal-expansion effect through the temperature-dependent equilibrium volume.
It does not include additional finite-temperature DOS broadening from random
thermal atomic displacements. The curves from 0 to 800~K are close to each
other because the static Al DOS changes only weakly over the thermally expanded
volume range. The inset magnifies the region near the Fermi level, where the
Fermi-Dirac occupation broadening produces the electronic entropy. The computed
$S_{\mathrm{ele}}$ increases almost linearly over the plotted range, and the
corresponding $-TS_{\mathrm{ele}}$ term becomes increasingly negative. The
purpose of plotting $-TS_{\mathrm{ele}}$ is to show the free energy
lowering caused by the electronic entropy, whereas $S_{\mathrm{ele}}$ itself
shows the entropy magnitude in units of $k_B$/atom. At 800~K,
$S_{\mathrm{ele}}=0.09645\,k_B$/atom and
$-TS_{\mathrm{ele}}=-6.649$~meV/atom. The full electronic correction entering
the free energy is smaller in magnitude because the electronic internal-energy
change partly compensates the entropy term, giving
$\Delta F_{\mathrm{ele}}\simeq -3.315$~meV/atom at 800~K after the low-temperature
intercept correction.

\section{Computational Formulae}
\label{app:computational_formulae}

The Hellmann-Feynman force on atom $i$ is evaluated as
\begin{equation}
\vec{F}_i = -\frac{\partial E_\text{tot}}{\partial \vec{R}_i},
\end{equation}
where $\vec{F}_i$ is the force on atom $i$, $E_\text{tot}$ is the total energy,
and $\vec{R}_i$ is the atomic position. In DFPT, the interatomic force
constant (IFC) matrix is obtained from the second derivative of the total
energy,
\begin{equation}
\Phi_{\alpha\beta}(i,j) =
\frac{\partial^2 E_\text{tot}}{\partial u_{i,\alpha} \partial u_{j,\beta}},
\end{equation}
where $\Phi_{\alpha\beta}(i,j)$ is the IFC element coupling Cartesian
directions $\alpha$ and $\beta$ of atoms $i$ and $j$, and $u_{i,\alpha}$
denotes the displacement of atom $i$ along Cartesian direction $\alpha$. The
IFCs are Fourier-transformed to construct the dynamical matrix
\begin{equation}
D_{\alpha\beta}(i,j|\vec{q}) =
\frac{1}{\sqrt{M_i M_j}} \sum_{\vec{R}}
\Phi_{\alpha\beta}(i,0; j,\vec{R})
e^{i\vec{q}\cdot\vec{R}},
\end{equation}
where $D_{\alpha\beta}(i,j|\vec q)$ is the dynamical-matrix element at phonon
wave vector $\vec q$, $M_i$ and $M_j$ are atomic masses, and $\vec{R}$ is the
lattice vector connecting periodic images. Phonon frequencies and eigenvectors
are then obtained from
\begin{equation}
\sum_{j,\beta} D_{\alpha\beta}(i,j|\vec{q})
e_{j,\beta}^n(\vec{q}) =
\omega_n^2(\vec{q}) e_{i,\alpha}^n(\vec{q}),
\end{equation}
where $e_{i,\alpha}^n(\vec q)$ is the phonon eigenvector component for branch
$n$ and atom $i$, and $\omega_n(\vec q)$ is the corresponding phonon
frequency.
The same eigenvalue problem can be written equivalently as
\begin{equation}
\det\!\left|D(\vec{q}) - \omega^2 I\right| = 0.
\end{equation}
Here, $I$ is the identity matrix and $\det$ denotes the determinant.
The phonon density of states (PDOS) is
\begin{equation}
g(\omega) = \frac{1}{N_q} \sum_{\vec{q}} \sum_n
\delta[\omega - \omega_n(\vec{q})],
\end{equation}
and the zero-point energy is
\begin{equation}
E_\text{ZPE} = \frac{1}{2N_q} \sum_{\vec{q}} \sum_n
\hbar\omega_n(\vec{q}),
\end{equation}
where $g(\omega)$ is the PDOS, $\delta$ is the Dirac delta function,
$E_\text{ZPE}$ is the zero-point energy, and $N_q$ is the total number of
sampled $q$ points.

\section*{Acknowledgements}

This work is financially supported by the National Natural Science Foundation
of China (No. 12074382, 11474285). We are grateful to the staff of the Hefei
Branch of Supercomputing Center of Chinese Academy of Sciences, and the Hefei
Advanced Computing Center for support of supercomputing facilities. We would
like to thank the crew of the Center for Computational Materials Science,
Institute for Materials Research of Tohoku University, and the supercomputer
resources through the HPCI System Research Project (hp200246). We also thank
Yan Gong for helpful discussions.

\section*{DATA AVAILABILITY}

The data that support the findings of this article are available within this
article and the Science Data Bank\cite{scidb}.

\end{document}